\documentclass{iopart}
\usepackage{graphicx}
\usepackage{amssymb}
\usepackage{setstack}
\usepackage{color}

\newcommand{\ket}[1]{\left|{#1}\right\rangle}
\newcommand{\bra}[1]{\left\langle{#1}\right|}

\begin{document}

\title{Correlated hopping of bosonic atoms induced by optical lattices}

\date{\today}

\author{Mar{\'\i}a Eckholt}
\address{Max-Planck-Institut f\"ur Quantenoptik,
  Hans-Kopfermann-Str.~1, Garching, D-85478, Germany}
\ead{maria.eckholt@mpq.mpg.de}

\author{Juan Jos\'e Garc{\'\i}a-Ripoll }
\address{Instituto de F\'{\i}sica
  Fundamental, CSIC, c/Serrano 113b, Madrid, E-28006, Spain.}

\begin{abstract}
  In this work we analyze a particular setup with ultracold atoms trapped
  in state-dependent lattices. We show that any asymmetry in the contact
  interaction translates into one of two classes of correlated
  hopping. After deriving the effective lattice Hamiltonian for the atoms,
  we obtain analytically and numerically the different phases and quantum
  phase transitions. We find for weak correlated hopping both Mott
  insulators and charge density waves, while for stronger correlated
  hopping the system transitions into a pair superfluid. We demonstrate
  that this phase exists for a wide range of interaction asymmetries and
  has interesting correlation properties that differentiate it from an
  ordinary atomic Bose-Einstein condensate.
\end{abstract}

 \submitto{\NJP}

\maketitle

\clearpage

\section{Introduction}

\begin{figure}
  \center{\includegraphics[width=0.7\linewidth]{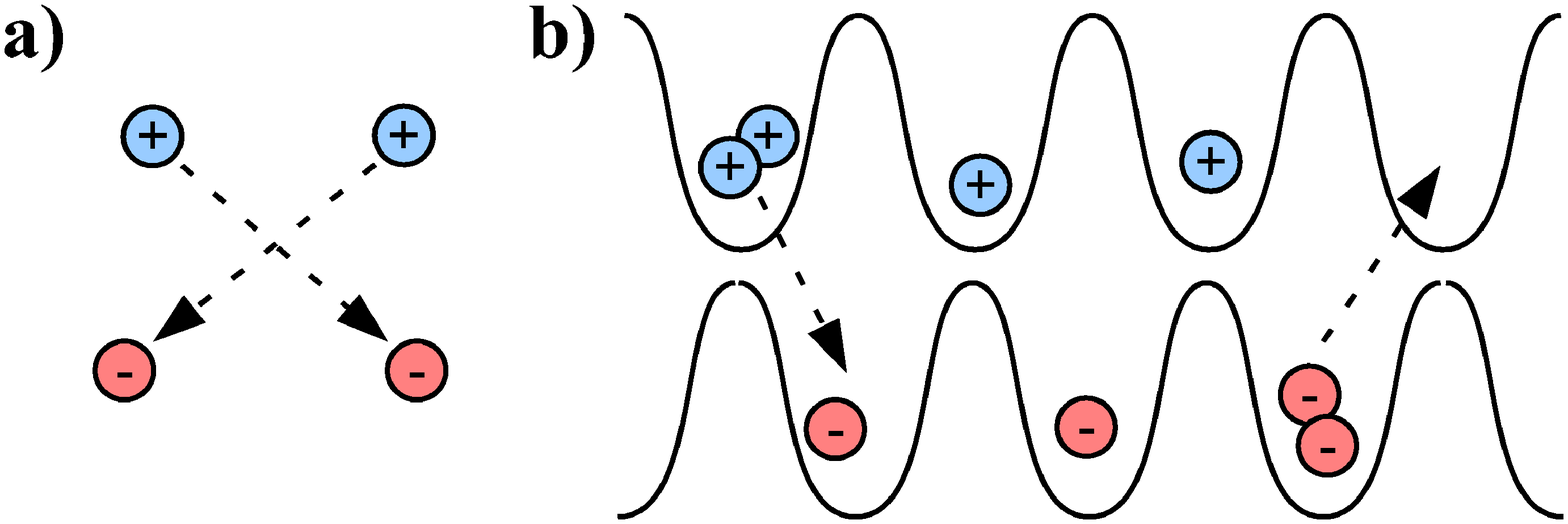}}
  \caption{(a) Two atoms in state $\ket{+}$ collide and flip state in free
    space. (b) When these atoms are confined in a state-dependent optical
    lattice, their change of state must be accompanied by a tunneling
    event to a neighboring site. This leads to correlated hopping of atoms
    through the lattice.}
  \label{fig:collisions}
\end{figure}

Ultracold neutral atoms are a wonderful tool to study many-body
physics and strong correlation effects. Since the achievement of
Bose-Einstein condensation in alkali atoms \cite{anderson95, davis95,
  bradley95}, we have been witnesses of two major breakthroughs. One
is the cooling of fermionic atoms and the study of Cooper pairing and
the BCS to BEC transition \cite{regal04, zwierlein04, bourdel04}. The
other one is the implementation of lattice Hamiltonians using bosonic
\cite{greiner02,jaksch98} and fermionic atoms in off-resonance optical
lattices. Contemporary and supported by these experimental
achievements, a plethora of theoretical papers has consolidated
ultracold atoms as an ideal system for quantum simulations. The goal
is two-fold: AMO is now capable of implementing known Hamiltonians
which could describe real systems in Condensed Matter Physics, such as
Hubbard models \cite{jaksch98} and spin Hamiltonians \cite{duan03,
  garciaripoll03}; but it is also possible to study new physical
effects, such as the quantum Hall effect with bosons \cite{paredes01}
and lattice gauge theories \cite{jaksch03, osterloh05}.

In this work we deepen and expand the ideas presented in \cite{eckholt08},
where we introduced a novel mechanism for pairing based on
transport--inducing collisions. As illustrated in
Fig.~\ref{fig:collisions}, when atoms collide they can mutate their
internal state. If the atoms are placed in a state-dependent optical
lattice, whenever such a collision happens the pair of atoms must tunnel to
a different site associated to their new state. For deep enough lattices,
as in the Mott insulator experiments \cite{greiner02}, this coordinated
jump of pairs of particles can be the dominant process and the ensemble may
become a superfluid of pairs.

The focus of this work is to develop these ideas for all possible
interaction asymmetries that can happen in experiments with bosonic atoms
in two internal states, where the scattering lengths among different states
can be different, $(g_{\uparrow\uparrow} \neq g_{\downarrow\downarrow})$
\cite{matthews99}. We want to understand the different types of correlated
hopping that appear when we move beyond the limited type of interactions
considered in Ref.~\cite{eckholt08}. We will better study the resulting
phases and phase transitions, using different analytical and numerical
tools that describe the many body states, and evolving from few particles
to more realistic simulations. The main results will be to show that when
one considers more general asymmetries, a new type of correlated hopping
appears, but that both the previous \cite{eckholt08} and the newly found
two--body terms cooperate in creating a superfluid state with pair
correlations. Indeed, this new state will also be shown to be more than
just a condensate of pairs, based on analytical and numerical studies.

Correlated hopping is not a new idea. It appears naturally in fermionic
tight-binding models, where it has been used to describe mixed valence
solids \cite{foglio79} and, given that they are able to mimic the
attractive interactions between electrons, also high-$T_{c}$
superconductors \cite{karnaukhov94, karnaukhov95, arrachea94, arrachea94b,
  deboer95, vidal01}. In most of these works, the correlated hopping
appears in the form $n_ia^{\dagger}_j a_k,$ indicating that the environment
can influence the motion of a particle. This would seem substantially
different from correlated motion of pairs of fermions. Nevertheless, even
this more elaborate form of correlated hopping has been shown to lead to
the formation of bound electron pairs \cite{arrachea94b, arrachea94} and it
has been put forward as a possible explanation for high $T_c$
superconductivity \cite{hirsch89, marsiglio90}.

This work is organized as follows. In Sec.~\ref{sec:themodel} we introduce
our model of correlated hopping (\ref{eq:model}), qualitatively discussing
how it originates and what are the quantum phases that we expect it to
develop. We present a possible implementation of this model which is based
on optical superlattices and atoms with asymmetric
interactions. Sec.~\ref{sec:exact} includes exact diagonalizations for a
small number of atoms and sites. These calculations reveal the existence of
insulating and coherent regimes, as well as pairing, and will be the basis
for later analysis. In Sec.~\ref{sec:analytics} we study the many-body
physics of larger lattices with correlated hopping, using a variety of
techniques: starting with insulating regime and following with the
implementation of perturbation theory and the quantum rotor model. These
methods suggest a number of possible phases, including a Mott insulator, a
pair superfluid, a normal superfluid and a charge density wave state, and
we estimate the parameters for which these phases appear. In
Sec.~\ref{sec:numerics} we develop two numerical methods to study our
system, a Gutzwiller ansatz and an infinite Matrix Product State
method. With these simulations we confirm the above mentioned phases and
locate the quantum phase transitions, which are found to be of second
order. Finally, in Sec.~\ref{sec:detection} we suggest some currently
available experimental methods to detect and characterize these phases.

\section{Correlated hopping model}
\label{sec:themodel}

\begin{figure}
  \includegraphics[width=\linewidth]{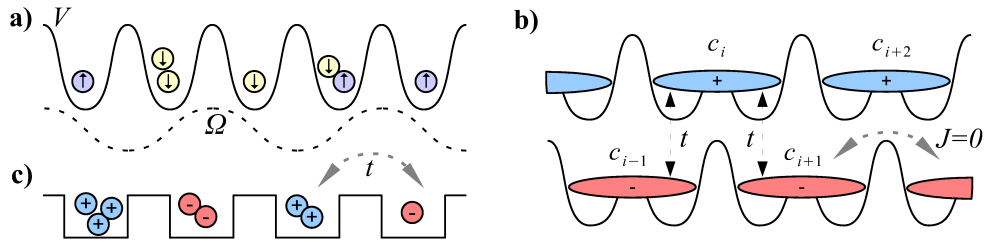}
  \caption{(a) Atoms in two internal states $\uparrow$ and $\downarrow$
    are trapped in an optical lattice and coupled by a Raman laser with
    Rabi frequency $\Omega$. (b) That setup is equivalent to two displaced
    superlattices for the dressed states $\ket{\pm} \sim \ket{\uparrow}\pm
    \ket{\downarrow}$. (c) When asymmetric contact interactions are
    considered and the hopping between superlattice cells neglected,
    the whole system behaves effectively as a $1$D array of alternating
    $\ket{+}$ and $\ket{-}$ sites, with transitions of tunneling amplitude
    $t$ between them.}
  \label{fig:superlattice}
\end{figure}

We suggested in Ref.~\cite{eckholt08} that the combination of atomic
collisions with optical superlattices can be used to induce correlated
hopping. The basic idea is shown in Fig.  \ref{fig:superlattice}b, where
atoms are trapped in two orthogonal states called $(+)$ and $(-).$ The
interaction terms change the state of the atoms, forcing them to hop to a
different superlattice every time they collide. In this sense, interactions
are responsible for transport.

In this section we introduce the most general model of correlated hopping
that can be produced by such means with two--states bosons. This model is
presented in the following subsection, where we explain qualitatively the
role of each Hamiltonian term. Later on, in
Sec.~\ref{sec:atomic-derivation}, we establish the connection between the
parameters of this model and the underlying atomic model. This is the
foundation for the subsequent analytical and numerical studies.

\subsection{Lattice Hamiltonian}

In this work we study the ground state properties of a very general
Hamiltonian that contains different kinds of correlated hopping.
More precisely, the model will be
\begin{eqnarray}
  H & = & \sum_i\bigg\{
    \frac{U}{4} :(n_i + n_{i+1})^2:+~V~n_i n_{i+1} -
    t~(c_i^{\dagger 2}c_{i+1}^2 + \mathrm{H.c.})\nonumber \\
    && ~~~~~~~ - j
    [(n_i - 1) c^\dagger_i(c_{i-1}+ c_{i+1}) + \mathrm{H.c.}]\bigg\}.
  \label{eq:model}
\end{eqnarray}
Here, $c_i^\dagger$ and $c_i$ are bosonic operators that create and
anihilate atoms according to the site numbering from
Fig.~\ref{fig:superlattice}b-c, and the colons $:A_iB_j:$ denote normal
ordering of operators $A_i$ and $B_j.$

Let us qualitatively explain the roles of the different terms in
Eq.~(\ref{eq:model}). The first and second terms, $U$ and $V,$ are related
to on-site and next-neighbor interactions. When these terms are dominant,
we expect the atoms in the lattice form an insulator.  Such a phase is
characterized by atoms being completely localized to lattice sites, having
well-defined occupation numbers, the absence of macroscopic coherence and a
gapped energy spectrum. Whether this insulating state is itself dominated
by strong on--site interactions $U$ or by nearest--neighbor
repulsion/attraction $V$ will decide whether it presents an uniform
density, a Mott insulator (MI), or a periodic density pattern, a charge
density wave (CDW), respectively.

The third term is the key feature of our model. It describes the tunneling
of pairs between neighboring lattice sites, with amplitude $t.$ Given
$U,V,j=0,$ we expect the atoms to travel along the lattice in pairs forming
what we call a pair superfluid (PSF). These pairs will be completely
delocalized, establishing long range coherence along the lattice. The
observable $\langle a^2\rangle$ would be the figure of merit describing
this kind of delocalization, while a vanishing $\langle a\rangle$ indicates
the absence of the single-particle correlations appearing in a normal
superfluid.  Furthermore, we expect this phase to have a critical velocity,
similar to that of an atomic condensate, and the energy spectrum should be
gapless.

Unlike in Ref.~\cite{eckholt08}, when one considers the most general kind
of atomic interaction, a second kind of correlated hopping appears,
described by the last term in Eq.~(\ref{eq:model}). Here, individual atoms
will hop only if there is already a particle in the site they go to
($c_i^\dagger c_j (n_i-1)$) or leave at least a particle behind ($(n_i-1)
c_j c_i^\dagger$). One might be induced to think that this term is
equivalent to single--particle hopping with a strength that depends on the
average density, thus giving rise to a single-particle superfluid (SF)
phase. However, this does not seem to be the case. We will show that the
correlated hopping $j$ generates a mixed phase which contains features of
both the ordinary BEC and the PSF created by $t.$

\subsection{Relation to atomic parameters}
\label{sec:atomic-derivation}

We now establish the relation between the model in
Eq.~(\ref{eq:model}) and the dynamics of atoms in an optical
superlattice. The actual setup we have in mind is shown in
Fig.~(\ref{fig:superlattice})a-b and described in more detail in
\ref{sec:trapping}. It consists on a three--dimensional lattice that
is strongly confining along the $Y$ and $Z$ directions, creating
isolated tubes. On top of this, we create an optical superlattice
acting along the $X$ direction \cite{eckholt08}. This superlattice
traps atoms in the dressed states $\ket{+}$ and $\ket{-},$ while the
atomic interaction is diagonal in the basis of bare states
$\ket{\uparrow}$ and $\ket{\downarrow}.$ The interaction will be
described by a contact potential and parameterized by some real
constants $g_{\alpha\beta}$
\begin{eqnarray}
  H_{\mathrm{int}} &=& \sum_{\alpha,\beta = \uparrow,\downarrow}
  \frac{g_{\alpha\beta}}{2}
  \int \psi_{\alpha}^\dagger(x) \psi_{\beta}^\dagger(x)
  \psi_{\beta}(x) \psi_{\alpha}(x) dx \label{eq:interaction1} \\
  &=&  \sum_{\alpha,\beta = \uparrow,\downarrow}
  \frac{g_{\alpha\beta}}{2} \int {: \rho_\alpha(x) \rho_\beta(x) :}\, dx
  \nonumber
\end{eqnarray}
These interaction constants are functions of the s--wave
one--dimensional scattering lengths between different species
$g_{\alpha\beta} = 4\pi\hbar^2 a^{(1D)}_{\alpha\beta}/m.$ In general,
the interaction strengths among different atomic components are
different from each other, a situation that can be enhanced with
Feshbach resonances. We will use a parameterization
\begin{eqnarray}
  g_{\uparrow\downarrow} = g_0 + g_1  = g_{\downarrow\uparrow},\quad
  g_{\uparrow\uparrow} = g_0 + g_2,\quad
  g_{\downarrow\downarrow} = g_0 - g_2,
\end{eqnarray}
that makes the symmetries more explicit
\begin{eqnarray}
  H_{\mathrm{int}} &=&
  \frac{g_0}{2}\int {:(\rho_\uparrow(x) +\rho_\downarrow(x))^2 :}\, dx
  +g_1\int {:\rho_\uparrow(x)\rho_\downarrow(x):}\, dx\nonumber \\
  & & + \frac{g_2}{2}\int {:\rho_\uparrow(x)^2 -\rho_\downarrow(x)^2 :}\, dx\label{eq:interaction2}
\end{eqnarray}
The total Hamiltonian combines the previous interaction with the kinetic
energy and the trapping potential for one particle
\begin{equation}
  H_{\mathrm{1}} = \sum_{s=\pm}\int \psi^\dagger_s(x)\left[
    -\frac{\hbar^2}{2m}\nabla^2 + V_s(x)\right]\psi_s(x) dx,
\end{equation}
which is written in a different basis
\begin{equation}
  \psi_\uparrow(x) = \frac{1}{\sqrt{2}} (\psi_+ + \psi_-),\quad
  \psi_\downarrow(x) = \frac{1}{\sqrt{2}} (\psi_+ - \psi_-).
  \label{eq:dressed}
\end{equation}
Since the superlattice potential $V_\pm(x)$ is the dominant term, we
may approximate the bosonic fields as linear combinations of the
Wannier modes in this superlattice and in the dressed state basis, a
process detailed in \ref{sec:derivation}. Note that out of all the
terms in the interaction Hamiltonian (\ref{eq:interaction2}), only
the first one is insensitive to the state of the atoms. This is
important because the asymmetries $g_1$ and $g_2,$ when expressed in
the dressed basis, produce terms that change the state of the atoms
during a collision. Once we introduce the effective interaction
constants in the lattice
\begin{equation}
  \label{int-constants}
  U_i = g_i \int|w(x)|^4 dx\quad\quad\mathrm{for}~i=0,1,2
\end{equation}
where $w(x)$ is the single site Wannier wavefunction, we arrive to
the effective Hamiltonian in Eq.~(\ref{eq:model}) with parameters
$U, V, t$ and $j$ that relate to the microscopic model as follows
\begin{equation}
\label{eq:paramet}
U = \frac{2U_0+U_1}{4},~~V = -\frac{U_1}{8}, ~~t =
\frac{U_1}{16}~~\mathrm{and}~~j = \frac{U_2}{8}.
\end{equation}
Unlike in the specific case of Ref.~\cite{eckholt08}, the most
general situation contains not only two-body correlated hopping $t,$
but also the terms proportional to $j.$

\section{Preliminary analysis}
\label{sec:exact}

In this section we study the eigenstates of Hamiltonian
(\ref{eq:model}) for systems that we can diagonalize exactly. The
goals are to characterize the effect of the different interaction
and hopping terms, as well as to understand the structure of the
ground state wavefunction. Although we are limited to a small number
of particles, the following examples provide enough evidence of the
roles of correlated hopping, nearest neighbor repulsion and the
utility of different correlators to characterize the states.

\subsection{A two-sites example}

Let us take the simplest interesting case: four particles in two sites. We
write the Hamiltonian in the basis $\{\ket{40}, \ket{22}, \ket{04},
\ket{31}, \ket{13}\},$ where the notation $\ket{n_1n_2}$ stands for $n_1$
particles in the first site and $n_2$ in the second and we restrict to $n_1
+ n_2=4,$
\begin{equation}\label{eq:Hfor4in2}
\fl~~~~~
    H_{\mathrm{4/2}} = \left(
      \begin{array}{ccccc}
        0         & -4\sqrt{6}t & 0           & -12j        & 0 \\
        -4\sqrt{6}t &  8V    & -4\sqrt{6}t & -6\sqrt{6}j & -6\sqrt{6}j \\
        0           & -4\sqrt{6}t & 0         & 0           & -12j \\
        -12j        & -6\sqrt{6}j & 0           & 6V    & -12t \\
        0           & -6\sqrt{6}j & -12j        & -12t        & 6V \\
      \end{array}
    \right) + 6 U.
\end{equation}
Notice that in this particular case, $U$ gives rise to a global
energy shift and does not affect the different eigenstates. This is
consistent with later studies where we will see that on-site
interactions just add a global, density dependent contribution to
the energy. To better understand the role of the remaining terms, we
will consider separately three limiting cases, two of superfluid
nature and an insulating one.

\paragraph{Limit $j\neq 0,$ $t=V=0,$ single-particle delocalization.}

In this case we take for simplicity $V=0$ and diagonalize Eq.
\ref{eq:Hfor4in2}, finding the normalized ground state
\begin{equation}
\label{eq:LowEn-j}
    \ket{\psi_{0,t=0}} = \frac{1}{4}\ket{40} +
    \frac{1}{2}\sqrt{\frac{3}{2}}\ket{22}
    + \frac{1}{4}\ket{04} + \frac{1}{2}\ket{31} +
    \frac{1}{2}\ket{13}.
\end{equation}
Note that this state is exactly a BEC of $4$ particles spread over
two sites
\begin{equation}
\label{eq:BEC-j}
    \ket{\psi_{BEC}(4)} = \frac{1}{\sqrt{4!}}\left(\frac{1}{\sqrt{2}} \sum_{i=1}^2
    c_i^\dagger \right)^4 \ket{00}.
\end{equation}
This suggests that, at least in this small example, the correlated hopping
proportional to $j$ is equivalent to the single-particle hopping in the
ordinary Bose-Hubbard model, giving rise to the delocalization of
individual particles. However, as it will become evident later on, for
larger systems and more particles this interpretation is wrong.

\paragraph{Limit $j=0,$ $t \gg |V|,$ pair delocalization.}

In the presence of two particle hopping, the lowest energy state has the
form
\begin{eqnarray}\label{eq:LowEn-t}
    \ket{\psi_{0,j=0}} = c_{40}(t,V)\ket{40} + c_{22}(t,V)\ket{22} +
    c_{04}(t,V)\ket{04}
\end{eqnarray}
with coefficients
\begin{eqnarray*}
  c_{22}(t,V) \propto -V+\sqrt{12t^2+V^2},\\
  c_{40}(t,V) = c_{04}(t,V) \propto \sqrt{6}t.
\end{eqnarray*}
In particular, for dominant pair hopping $t \gg |V|$ this is a state of
delocalized pairs
\begin{equation}\label{eq:4/2full}
\ket{\psi} = \frac{1}{2}\ket{40} + \frac{1}{\sqrt{2}}\ket{22} +
\frac{1}{2}\ket{04}.
\end{equation}
Observe that this wavefunction is not equivalent to what one would
na\"ively understand as a ``pair condensate'' from analogy with the
single-particle case
\begin{equation}
\ket{\psi} \neq \left(\sum_{i=1}^2 c_i^{\dagger
    2}\right)^{2} \ket{\mathrm{vac}}
\sim
\sqrt{\frac{3}{8}}\ket{40} + \frac{1}{2}\ket{22} +
\sqrt{\frac{3}{8}}\ket{04}.\label{eq:likeBEC}
\end{equation}
Instead the previous wavefunction is isomorphic to the BEC of two bosons
\begin{equation}\label{condensate2}
  \ket{\psi_{BEC}(2)} = \frac{1}{2}\ket{20} + \frac{1}{\sqrt{2}}\ket{11} +
  \frac{1}{2}\ket{02}
\end{equation}
under the replacement of each boson with two atoms. It is also
interesting to remark that $\psi_{BEC}(2)$ has larger
pair-correlations than the state in Eq.~(\ref{eq:likeBEC}).

\paragraph{Limit $|V|\to\infty,$ an insulator.}

Reusing the previous wavefunction (\ref{eq:LowEn-t}) and taking the
limit of dominant nearest-neighbor interaction $V,$ we obtain two
possible states. For strong repulsion $V\to +\infty,$ states
$\ket{40}$ and $\ket{04}$ are favored, forming a charge density wave
(CDW) with partial filling $\ket{\psi_{CDW}} \propto
\ket{40}+\ket{04}.$ On the other hand, for strong nearest-neighbor
attractions $V\to -\infty,$ the particles are evenly distributed
forming a Mott insulator $\ket{22}.$

\subsection{Superfluidity of pairs}\label{sec:SFpairs}

We have seen that four particles in a two-sites lattice recreate the exact
wavefunction of an ordinary BEC under the replacement of single bosons with
pairs. We can test this idea for slightly bigger lattices, diagonalizing
numerically the Hamiltonian which only contains the pair hopping term
($t\neq 0,~ V,U,j=0$). The resulting wavefunctions are compared side by
side with the BEC-like ansatz we mentioned. In the case of two particles we
get indeed the expected result
\begin{eqnarray*}
  \ket{\psi_2^{g.s.}} &=& \ket{\psi_2^{ideal}} \sim \sum_{i=1}^L a^{\dagger 2}_i
  \left|\mathrm{vac}\right\rangle,
\end{eqnarray*}
whereas for $4$ particles in $5$ sites
\begin{eqnarray*}
  \ket{\psi_4^{g.s.}} &=&
    c_1\big(\ket{40000} + \ket{04000} + \ket{00400} + \ket{00040} +
    \ket{00004}\big) \\
    && + c_2\big(\ket{22000} + \ket{02200} + \ket{00220} + \ket{00022} +
    \ket{20002}\big) \\
    && + c_3\big(\ket{20200} + \ket{20020} + \ket{02020} + \ket{02002} +
    \ket{00202}\big),
\end{eqnarray*}
we find a disagreement between the ideal case of a BEC-like state
with coefficients $c_1=1/5, c_2=c_3=\sqrt{2}/5,$ and the exact
diagonalization with $c_1 \sim 0.2735,$ $c_2 \sim 0.3073$ and $c_3
\sim 0.1754.$ We observe that when compared to the ideal BEC, our
paired state breaks the translational symmetry, revealing an
effective attraction between different pairs, that favors their
clustering.

\begin{figure}
  \centering
  \includegraphics[width=0.513\linewidth]{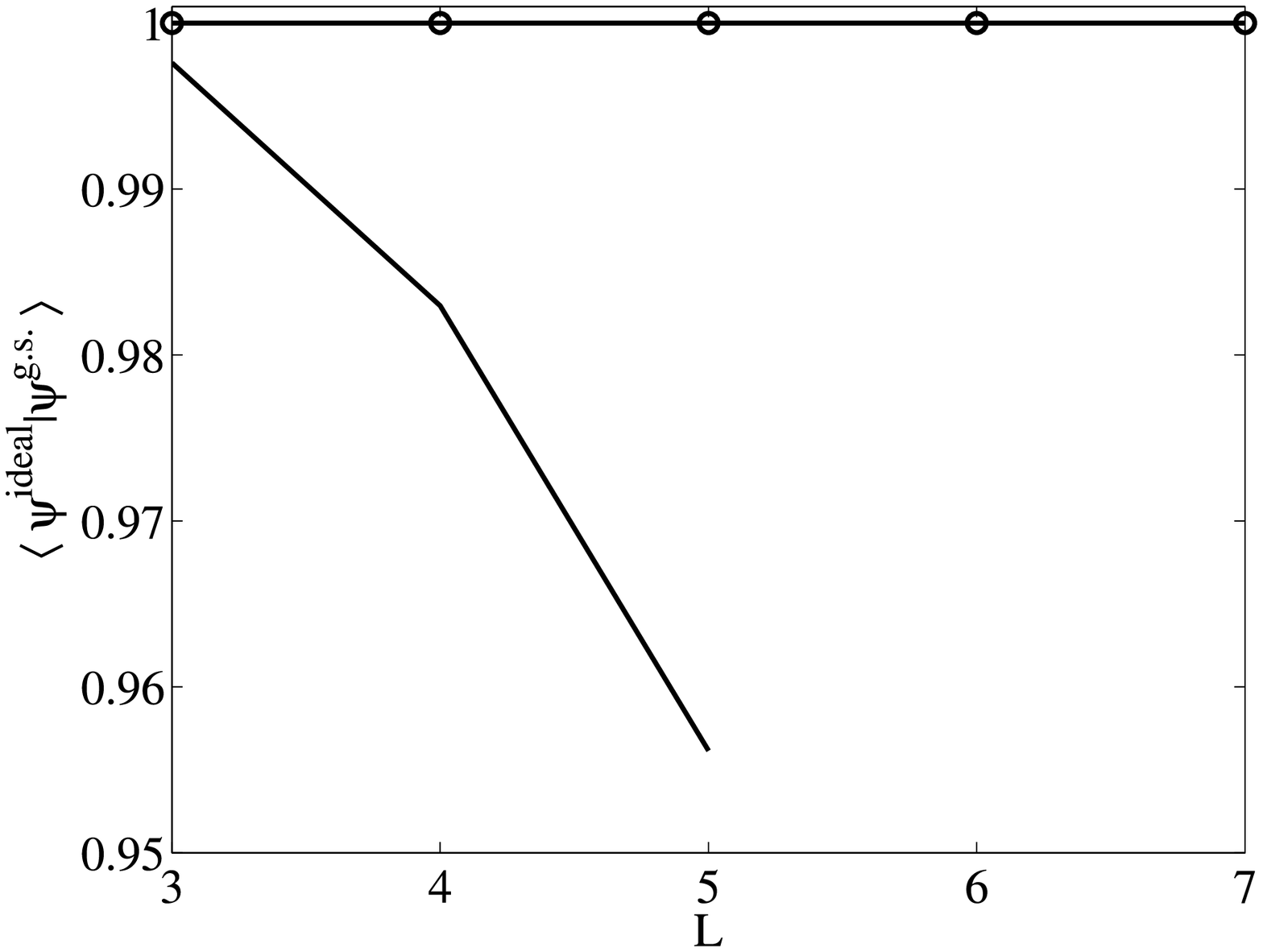}%
  \includegraphics[width=0.47\linewidth]{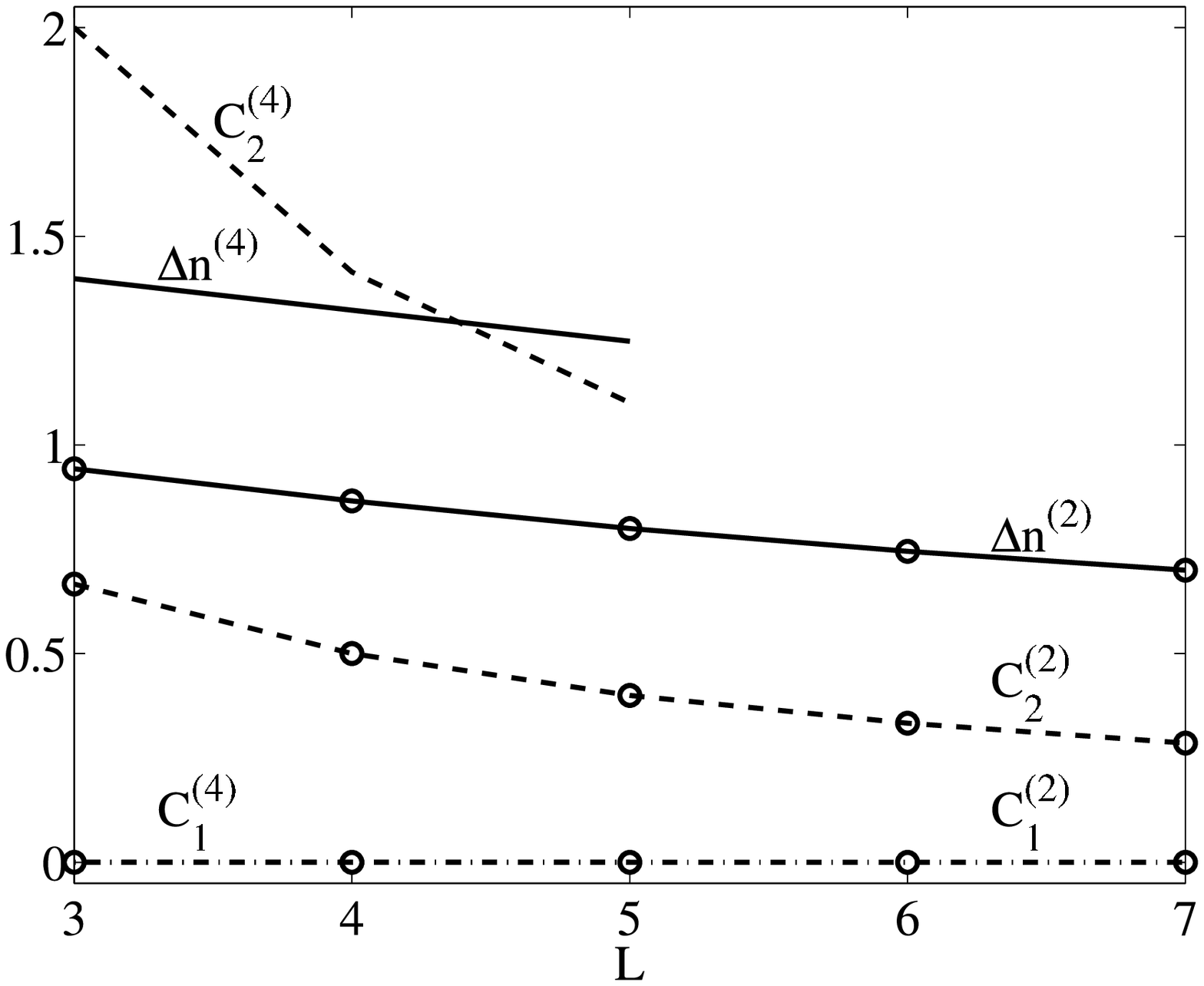}
  \caption{(left) Fidelity between the ground state of the Hamiltonian
  and the \emph{ideal} PSF for the case of 2 particles (line with circles) and 4
  (no markers). (right) Particle fluctuation (solid), one-body correlator (dot-dash)
  and two-body correlator (dash) for the case of 2 particles (circle) and 4
  (no markers) using the ground state of the Hamiltonian.}
  \label{fig:compPSF}
\end{figure}

In Fig.~\ref{fig:compPSF} we plot the projection between these
states, namely the solution of Eq. \ref{eq:model} with only $t\neq
0$ and the \emph{ideal} superfluid of pairs. In the nearby plot we
also analyze two relevant correlators that will be also used later
on in the manuscript, namely, a single-particle coherence
\begin{equation}
 C_\Delta^1 = \frac{1}{L}\sum_i\left \langle a^{\dagger}_{i+\Delta}
 a_i\right \rangle
\end{equation}
and the pair correlator
\begin{equation}
 C_\Delta^2 = \frac{1}{L}\sum_i\left \langle a^{\dagger 2}_{i+\Delta}
 a_i^2\right \rangle.
\end{equation}
As it is evident from the wavefunction and from the plots, there is
no single-particle coherence or delocalization because particles
move in pairs. Hence, $C^1_\Delta\sim \delta_{\Delta 0}.$ The other
correlator, $C^2_\Delta,$ which we identify with the delocalization
of pairs is rather large and it only decreases with increasing the
lattice size because the total pair density becomes smaller.

\section{Analytical methods}
\label{sec:analytics}

We now study the many-body physics of our model for a much larger
number of particles using exact analytical methods. We begin with
the regime in which the interaction terms $U$ and $V$ dominate,
obtaining the different insulator phases on the $j,t=0$ plane. Then,
using perturbation theory, we compute the phase boundaries of these
insulating regions for growing $j$ and $t.$ Finally, we study the
properties of the ground state and its excitations in the superfluid
phase, with $j=0$ and dominating $t,$ proving indeed that this
region describes a superfluid of pairs.

\subsection{No hopping limit: insulating phases}
\label{sec:NoHopp}

To analyze the phase diagram it is convenient to work in the
grand-canonical picture, in which the occupation is determined by the
chemical potential $\mu.$ In this picture the ground state is
determined by minimizing the free energy
\begin{equation}
  F = H - \mu N,
\end{equation}
where $N=\sum_k n_k$ is the \textit{total} number of particles,
including both states $\ket{+}$ and $\ket{-}.$ The free energy has a
very simple form in the absence of tunneling
\begin{eqnarray}
  F &=& \sum_k\left[ \frac{U}{4} : (n_k + n_{k+1})^2 :  + V n_k n_{k+1} -
  \mu n_k \right] \label{eq:interaction} \\
  &=& \sum_k\left[ \frac{U}{4} (n_k + n_{k+1})^2  + V n_k n_{k+1} -
  \left(\frac{2\mu + U}{4} \right) (n_k + n_{k+1}) \right]. \nonumber
\end{eqnarray}
This function is defined over positive occupation numbers $n_k \in
\{0,1,2,\ldots\}.$ A discrete minimization will determine the
different insulating phases and the regions where the system is
stable against collapse.

For a translational invariant system with periodic boundary
conditions, all solutions can be characterized as a function of two
integers ${\vec x}^t=(n,m),$ representing the occupations of even
$n_{2k}=n$ and odd sites $n_{2k+1}=m.$ The optimization begins by
noticing that the bond energy of two sites has a quadratic form
\begin{eqnarray}
  \varepsilon(\vec x) &=& {\vec x}^t
  \left(
    \begin{array}{cc}
      U/4 & (U + 2V)/4 \\
      (U + 2V)/4  & U/4
    \end{array}
  \right) \vec x -
  {\vec x}^t
  \left(
    \begin{array}{cc}
      (U + 2\mu)/4 \\
      (U + 2\mu)/4
    \end{array}
  \right) \nonumber\\
  &=& {\vec x}^t A \vec x - {\vec x}^t \vec v,
\end{eqnarray}
where physical solutions are in the sector with $n,m\geq 0.$ For these
occupation numbers to remain bounded, the bond energy
$\varepsilon(\vec x)$ has to increase as $n, m$ or both grow. This
gives us two conditions that need to be fulfilled to prevent
collapse. If these conditions are not met, the ground state will be an
accumulation of all atoms in the same site. In that case, the large
interaction energies and the many--body losses induced by the large
densities will cause the breakdown of our model and quite possibly of
the experimental setup.

The first stability condition is found by studying $\varepsilon(\vec x)$
along the boundaries of our domain ($n, m \geq 0$). Take for instance
$m=0,$ this gives a total energy $\varepsilon_B = (U/4)n^2 -
[(U+2\mu)/4]n.$ For this function to have a local minimum at finite $n,$ we
must impose
\begin{equation}\label{eq:restriction1}
  U > 0.
\end{equation}
The second condition comes from analyzing the interior of the
domain. For this, we take the only eigenvector of $A$ which lies in
the region of positive occupation numbers, $n=m=x/2.$ This line has
an energy $\varepsilon_+ = [(U+V)/4]x^2 - [(U+2\mu)/4]x$ and, to
have again a finite local minimum, we require
\begin{equation}\label{eq:restriction2}
V > -U.
\end{equation}

\begin{figure}[t]\label{fig:stability}
  \centering
  \includegraphics[width=0.4\linewidth]{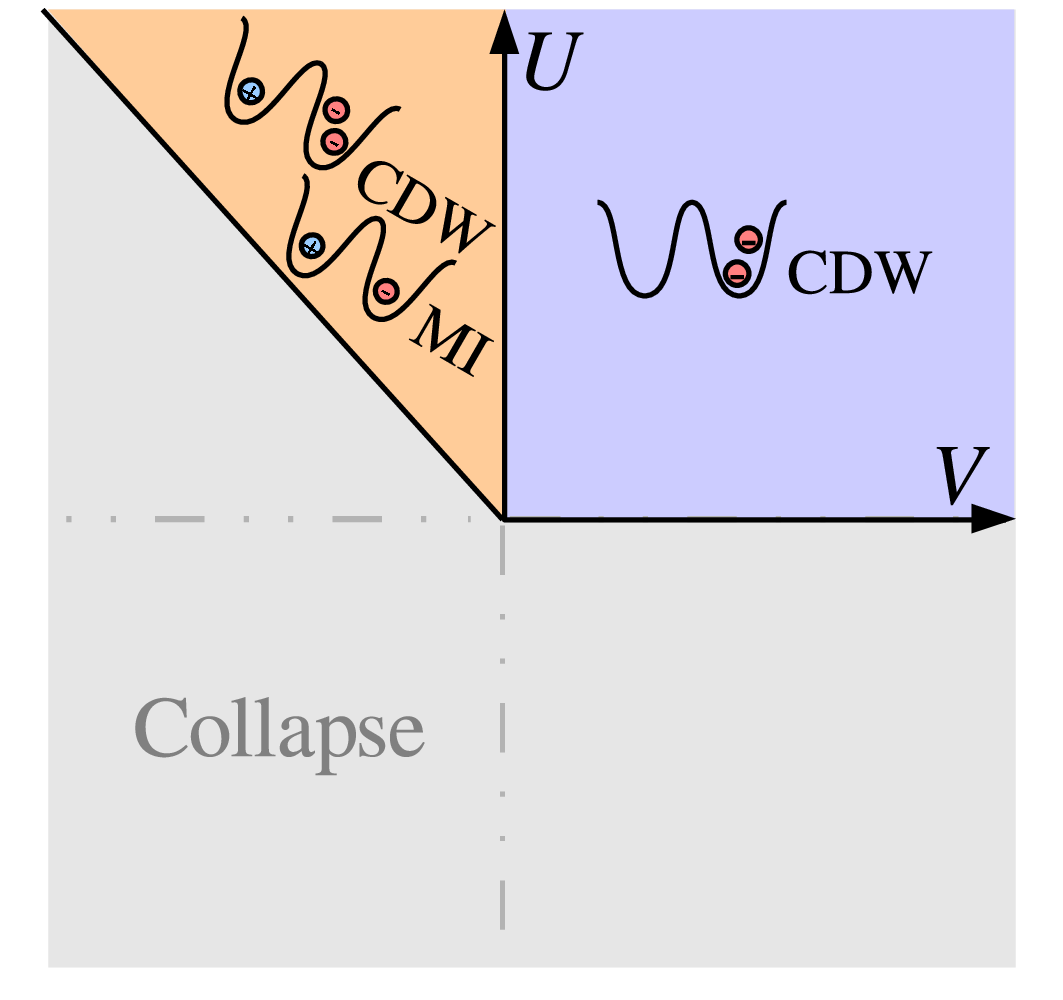}%
  \caption{Phase diagram for $U,V$: different regions of stability
    and insulating phases for $t,j=0.$}
\end{figure}

Given that Eq.~(\ref{eq:restriction1}) and
Eq.~(\ref{eq:restriction2}) are satisfied, the system is stable and
we have two possibilities to attain the minimum energy: either at
the boundaries, $n = 0$ or $m = 0,$ or right on the eigenvector of
$A.$ Inspecting $\varepsilon_B$ and $\varepsilon_+$ we conclude that
a positive value of $V$ will lead to the formation of charge density
waves (CDW) of filled sites alternating with empty sites
\begin{equation}
  V > 0 \, \Rightarrow \, n_{2k} = 0 ~\mbox{or}~ n_{2k+1} = 0.
\end{equation}
If $V\leq 0$ our energy functional will be convex and the minimum
energy state will be a Mott insulator with $n = m,$ when $n + m$ is
even, or a charge density wave with $n = m \pm 1,$ when $n + m$ is
odd. The actual choice between these two insulating phases is
obtained by computing the energy of both states
\begin{eqnarray}
  \epsilon(2n+1) &=& U (2n+1)^2/4 + V n(n+1) - (U+2\mu) (2n+1)/4\\
  \epsilon(2n) &=& U (2n)^2/4 + V n^2 - (U+2\mu) 2n/4.
\end{eqnarray}
Having $\epsilon(2n+1) - \epsilon(2n) = 0$ defines the value of
$\mu$ at which the state with $2n$ particles every two sites, a Mott
with $n$ particles, stops being the ground state and becomes more
favorable to acquire an extra particle to form a CDW. The boundaries
of these insulating phases for $t, j = 0$ are given by
\begin{eqnarray}
  \mu(2n\to 2n+1) = (U+V) 2n,\\
  \mu(2n-1\to 2n) = (U+V) 2n - U.
\end{eqnarray}
Thus summing up, for $\mu(2n-1\to 2n)\leq \mu \leq \mu(2n\to2n+1)$
the optimal occupation is $n$ particles per site, forming a Mott,
while for $\mu(2n\to2n+1)\leq \mu\leq \mu(2n+1\to 2n+2)$ the
occupation number is $2n+1$ particles spread over every two sites,
having a CDW. The results of this section are summarized in
Fig.~\ref{fig:stability}.

\subsection{Perturbation theory: insulator phase boundaries}
\label{sec:PertTeo}

The previous calculation can be improved using perturbation theory
for $t,j \ll U,V$ around the insulating phases, obtaining the phase
boundaries around the insulators as $t$ and $j$ are increased. This
is done applying standard perturbation theory up to second order on
both variables \cite{FreericksPRB96}, using as unperturbed
Hamiltonian the operator (\ref{eq:interaction}) and as perturbation
the kinetic energy term
\begin{equation}\label{eq:kinetic}
  W = \sum_i\left\{- \left[t~c_i^{\dagger 2}c_{i+1}^2 +
  j~(n_i - 1) c^\dagger_i(c_{i-1}+ c_{i+1})\right] + \mathrm{H.c.} \right\}.
\end{equation}

We start calculating analytically the ground state energies of the
first four insulating phases according to (\ref{eq:interaction}),
considering the perturbation $W$ up to second order in $j,t.$ For
the CDW with $n_i = 1$ and $n_{i+1} = 0$ this energy is obviously
zero
\begin{equation}
  E(L/2) = 0.
\end{equation}
For the MI with one particle per site we have virtual processes of
the correlated hopping $j,$ as environment-assisted hopping starts
being allowed in an uniformly filled lattice
\begin{eqnarray}
  E(L) = (U + 2V)L/2 - \frac{8j^2}{U-2V}L.
\end{eqnarray}
For the CDW with $n_i = 2$ and $n_{i+1} = 1,$ we find some doubly
occupied sites and contributions from the pair hopping $t$
\begin{eqnarray}
  E(L + L/2) = (3U + 4V)L/2 - \left(\frac{6t^2}{U} +
    \frac{24j^2}{U-6V} + \frac{32j^2}{U+2V}\right)L.
\end{eqnarray}
Finally, for the MI with two particles per site, a calculation
detailed in \cite{Eckholt09},
  \begin{eqnarray}
    E(2L) = (3U + 4V)L - \frac{24t^2 + 216j^2}{U-2V}L.
  \end{eqnarray}
Here $L$ is the total number of sites and all results presented in
this section are for the case $V < 0.$

\begin{figure}[t]
  \centering
  \includegraphics[width=0.47\linewidth]{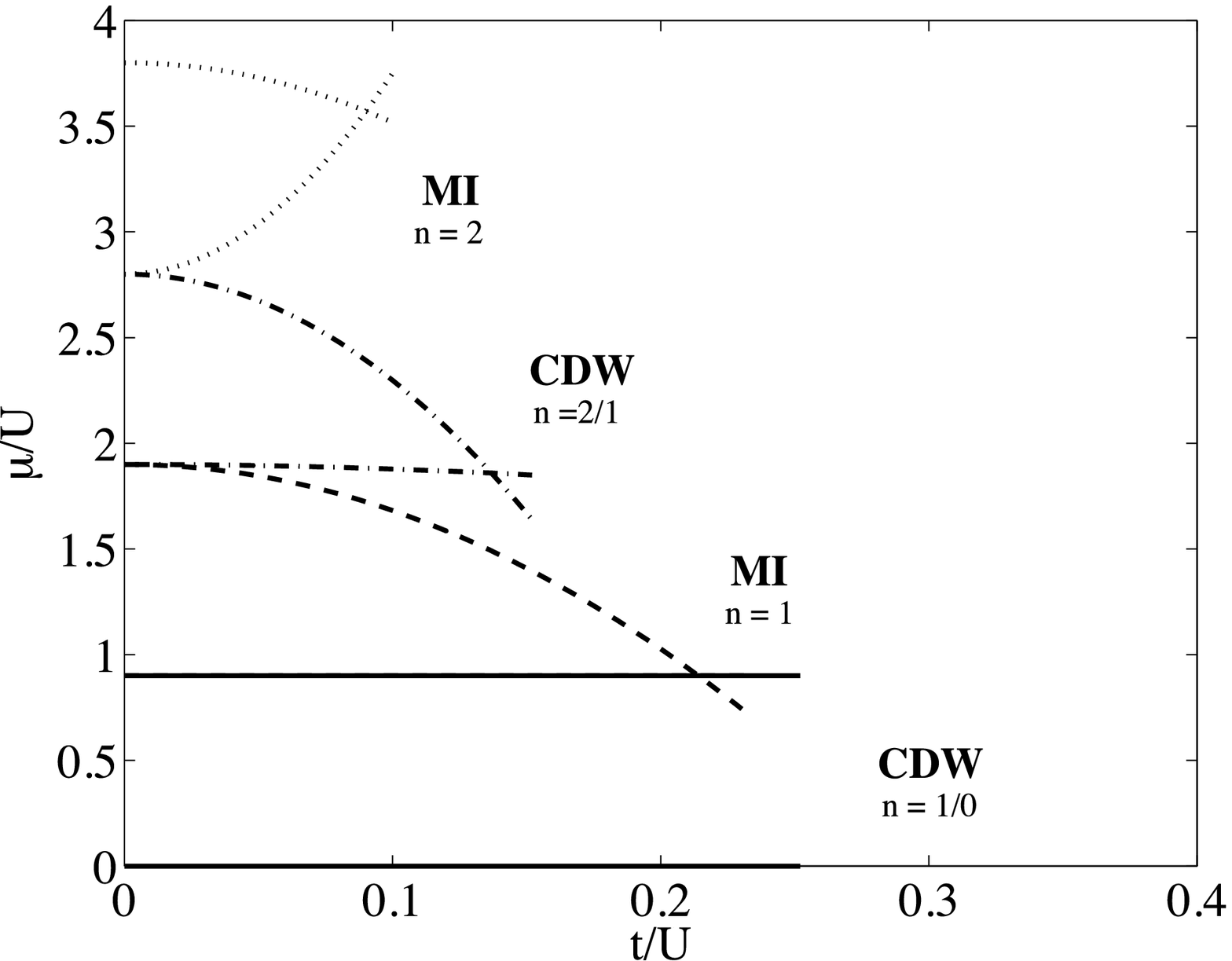}%
  \includegraphics[width=0.47\linewidth]{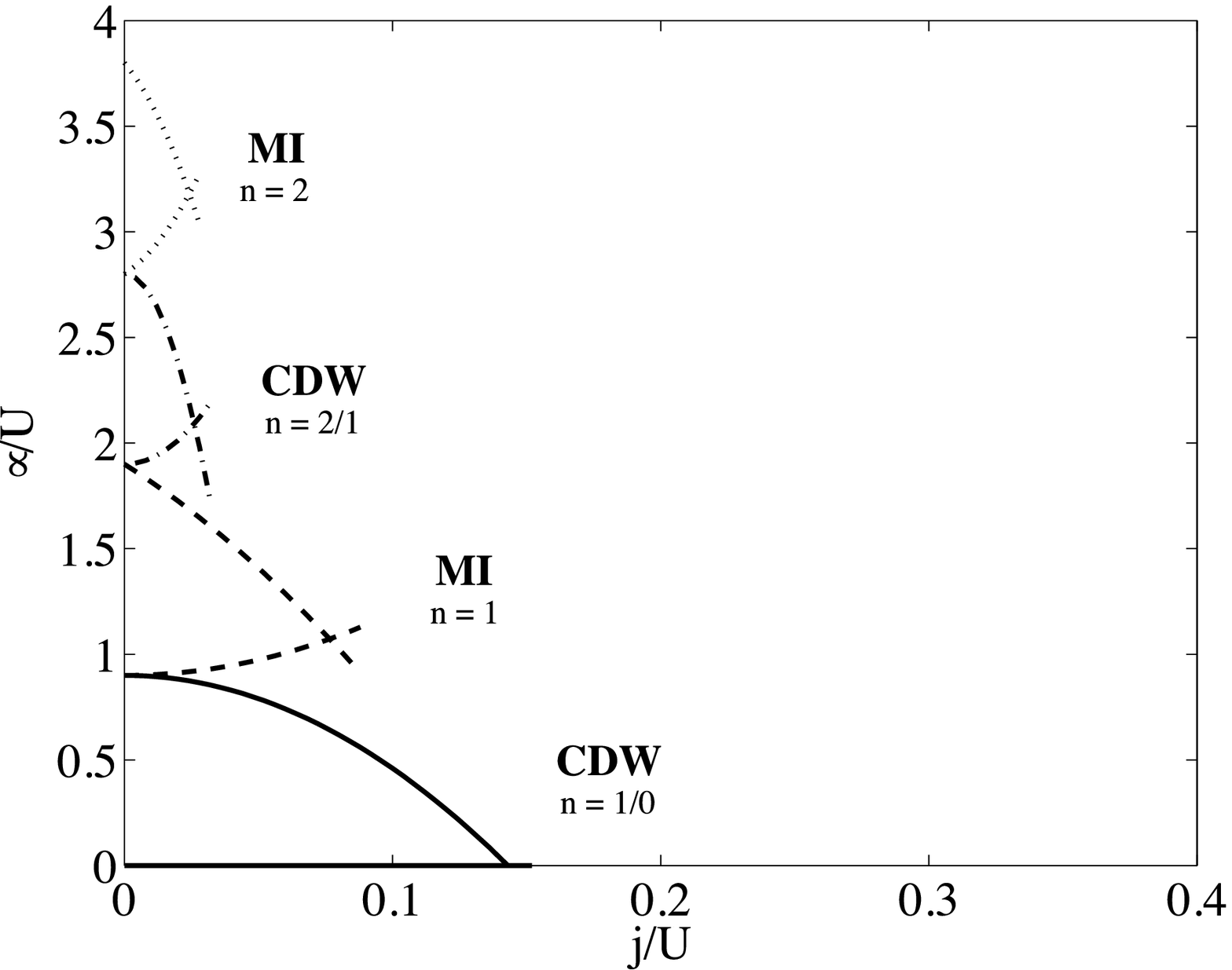}
  \caption{Phase diagrams for $\mu,$ $j,t\ll |U|$ and
    $V = -0.05.$ (left) Varying $t$ for $j = 0.$ (right) Varying
    $j$ for $t = 0.$ In both cases the lowest region is a
    CDW with alternating $0$ and $1$ particle occupation, followed
    upwards by a Mott of one particle per site, a CDW with $1$ and
    $2$ particles and the highest area a Mott of two particles.}
  \label{fig:PTlimits}
\end{figure}

At each value of $j,t,$ the boundary of an insulating phase with average
density $\bar n$ is given by the degeneracy condition with a compressible
state $E(\bar nL) = E(\bar nL \pm 1).$ Those points correspond to the
chemical potential at which a hole, $\mu_{h}(\bar nL) = E(\bar nL) - E(\bar
nL-1),$ or a particle, $\mu_{p}(\bar nL) = E(\bar nL+1) - E(\bar nL),$ can
be introduced We show here the lower and upper limits of the first four
insulating regions, corresponding to the CDW with $n_i = 1, n_{i+1} = 0$
  \begin{eqnarray}
    \mu_h(L/2) &=& 0 \nonumber\\
    \mu_p(L/2) &=& U + 2V + j^2(\frac{2}{V} - \frac{4}{U})
  \end{eqnarray}
the Mott with one particle per site
  \begin{eqnarray}
  \mu_h(L) &=& U + 2V + j^2\left(\frac{4}{U} - \frac{2}{V} -
      \frac{16}{U - 2V}\right) \\
  \mu_p(L) &=& E(L + 1) - E(L) \nonumber\\
    &=& 2U + 2V - 8j - \frac{4j^2}{U} + \frac{2j^2}{V} +
      \frac{8(j^2 - 6jt - 3t^2)}{U - 2V}
  \end{eqnarray}
the CDW with $n_i = 2, n_{i+1} = 1$
  \begin{eqnarray}
  \mu_h(L + L/2) &=& 2U + 2V + \frac{48t^2V}{U^2 - 2UV}\nonumber\\
      &&+ j^2\left(\frac{4}{U} - \frac{2}{V} + \frac{96}{U - 6V}
      + \frac{24}{U - 2V} + \frac{128}{U + 2V}\right) \\
  \mu_p(L + L/2) &=& E(L + L/2 + 1) - E(L + L/2) \nonumber\\
    &=& 3U + 4V + j^2\left(- \frac{108}{U} + \frac{54}{V}
      + \frac{96}{U - 6V} + \frac{64}{U + 2V}\right) \nonumber\\
      &&+ 6t^2\left(\frac{4}{U} - \frac{8}{3U - 2V}
      - \frac{4}{U - 2V} - \frac{8}{U - 6V}\right)
  \end{eqnarray}
and the MI with two particles per site
\begin{eqnarray}
  \mu_h(2L) &=& 3U + 4V + 8j + \frac{108j^2}{U}
  + 24t^2\left(\frac{2}{U - 6V} + \frac{2}{3U - 2V}\right)
  \nonumber\\
  &&- \frac{54j^2}{V} - \frac{24(35j^2 -
    3\sqrt{2}jt + 3t^2 )}{U - 2V} \\
  \mu_p(2L) &=& 4U + 4V - 6\sqrt{6}j - \frac{108j^2}{U}
  - \frac{48t^2}{U - 6V} - \frac{48t^2}{3U - 2V} \nonumber\\
  &&+ \frac{54j^2}{V} + \frac{32(57j^2 - 12\sqrt{6}jt -
    7t^2)}{3U - 6V}
\end{eqnarray}
The corresponding boundaries are plotted in Fig.~\ref{fig:PTlimits}.
For small hopping amplitude, they match the values that are found
later on with the numerical methods. But even for larger values,
this approximation anticipates that the lobes are significantly
larger for pair hopping $t$ than for the correlated hopping $j.$

\subsection{Phase model: analysis of the pair condensate}

So far we have studied the many-body physics around the limit of strong
interactions. However, the main goal of this work is to understand the
effect of correlated hopping and the creation of a pair superfluid. In
absence of a mean field theory, but still in the limit of dominant two-body
hopping $U,V \ll t,$ we can use the number-phase representation, introduced
in Ref.~\cite{garciaripoll04} for an ordinary BEC. Note, however, that the
model in Ref.~\cite{garciaripoll04} cannot be directly applied
here. Following that reference, one would assume a large number of
particles per site, $n_i > 1,$ and introduce the basis of phase states
$|\vec{\phi}\rangle$
\begin{equation}\label{firstPM}
    \langle \vec{n}|\vec{\phi} \rangle =
    (2\pi)^{-L/2}e^{i\vec{n}\cdot\vec{\phi}}.
\end{equation}
Using these states, one would then develop approximate representations for
the operators $a_i^2,$ $a_i^{\dagger 2}$ and $n_i,$ and diagonalize the
resulting Hamiltonian in the limit of weak interactions. But after a few
considerations one finds that the resulting phase model does not preserve
an important symmetry of our system: if $j=0$ particles can only move in
pairs and the parity of each site, $(-1)^{n_i},$ is a conserved quantity.

To describe correlated hopping we must use a basis of
states with fixed parity $\nu$
\begin{equation}\label{secondPM}
    \langle 2\vec{n} + \nu'|\vec{\phi} \rangle_\nu =
    (2\pi)^{-L/2}e^{i\vec{n}\cdot\vec{\phi}} \delta_{\nu\nu'},\quad\nu\in\{0,1\}
\end{equation}
which is $\nu=0$ for the ground state we are interested in. As mentioned
before, we now have to find expressions for the different operators,
$a_i^2,$ $a_i^{\dagger 2}$ and $n_i.$ We use the fact that our states will
have a density close to the average value $\bar n$ and approximate the
action of the operators over an arbitrary state as
\begin{eqnarray}
  n_i\ket{\vec{\phi}}
  &=& (-i2\partial_{\psi_i}-\bar{n})\ket{\vec{\phi}}, \\
  a_i^2\ket{\vec{\phi}}
  &=& \sqrt{\bar{n}(\bar{n}-1)}e^{-i\phi_i} \ket{\vec{\phi}}, \\
  a_i^{\dagger 2}\ket{\vec{\phi}}
  &=& \sqrt{(\bar{n}+1)(\bar{n}+2)}e^{i\phi_i}\ket{\vec{\phi}}.
\end{eqnarray}
Introducing the constant $\rho^2 =
\bar{n}(\bar{n}-1)(\bar{n}+1)(\bar{n}+2)$ our Hamiltonian becomes
similar to the quantum rotor model \cite{garciaripoll04}
\begin{eqnarray}\label{secondPM_H}
    H &=& \sum_{i=1}^L
    \left[2U \partial_{\phi_i}^2 + (2U + 4V)
    \partial_{\phi_i}\partial_{\phi_{i+1}}+
    2\rho^2t\cos\left(\phi_i - \phi_{i+1}\right)\right]\nonumber\\
    &&-(U + V)L\bar{n}^2.
\end{eqnarray}
For small $U$ and $V,$ the ground state of this model is
concentrated around $\phi_i-\phi_{i+1}=0.$ Expanding the Hamiltonian
up to second order in the phase fluctuations around this equilibrium
point, we obtain a model of coupled harmonic oscillators. This new
problem can be diagonalized using normal modes that are
characterized by a quasi-momentum $k=2\pi n/L,$ $n \in
[-(L-1)/2,(L-1)/2].$ The result is
\begin{equation}\label{PM_diag}
    H = \sum \hbar \omega_{k}\left(b_k^\dagger b_k + \frac{1}{2}\right) + E_0
\end{equation}
with normal frequencies
\begin{equation}\label{PM_freq}
    \omega_k =
    4\rho\sqrt{2Ut}\left|\sin(k/2)\right|\sqrt{1 +
    \left(1+\frac{2V}{U}\right)\cos\left(k\right)},
\end{equation}
and a global energy shift $E_0 = 4(U + V) N^2 - 4 L \bar n^2
(U+V) - 2 \rho^2 t L.$

It is evident from Eq.~(\ref{PM_freq}) that our derivation is only
self consistent for negative values of $V.$ Otherwise, when $V>0$
some of the frequencies become imaginary, signaling the existence
of an unbounded spectrum of modes with $|k| \ge \pi/4$ and that our
ansatz becomes a bad approximation of the ground state. This
strictly means that our choice $\phi_i=\phi_{i+1}$ only applies in
the case of attractive nearest neighbor interactions, $-U \leq V
\leq 0,$ as we know that this interaction cannot destabilize a
translational invariant solution such as the uniform Mott insulator.
However, it does not mean by itself that the whole system becomes
unstable for $V>0$ --- indeed, we will show numerically that it
remains essentially in a similar phase for all values of $V,$ but in
the case of $V>0$ the insulating phases are stable until values of
the hopping slightly higher as in the $V<0$ case.

If we focus on the regime of validity, we will find that the spectrum is
very similar to that of a condensate. At small momenta the dispersion
relation becomes linear, $\omega_k \propto v_g k,$ with sound velocity
$v_g=4\rho\sqrt{2Ut}/\hbar,$ while at larger energies the spectrum becomes
quadratic, corresponding to ``free'' excitations with some mass. This a
consequence of the similarity between our approximate model for the pairs
(\ref{secondPM_H}) and the phase model for a one-dimensional
condensate. However, we can go a step further and conclude that the
similarity extends also to the wavefunctions themselves, so that the state
of a pair superfluid can be obtained from that of an ordinary BEC by the
transformation $n \to 2n.$ This is indeed consistent with what we obtained
for the diagonalization of a two-particle state in the limit $j,U,V=0$ [See
Eq.~(\ref{condensate2})].

\section{Numerical methods}\label{sec:numerics}

The previous sections draw a rather complete picture of the possible
ground states in our model. In the limit of strong interactions we
find both uniform insulators and a breakdown of translational
invariance forming a CDW, while for dominant hopping we expect both
single-particle superfluidity and a new phase, a pair superfluid. We
now confirm these predictions using two different many-body
variational methods.

\subsection{Gutzwiller phase diagram}

\begin{figure}
  \centering
  \includegraphics[width=0.5\linewidth]{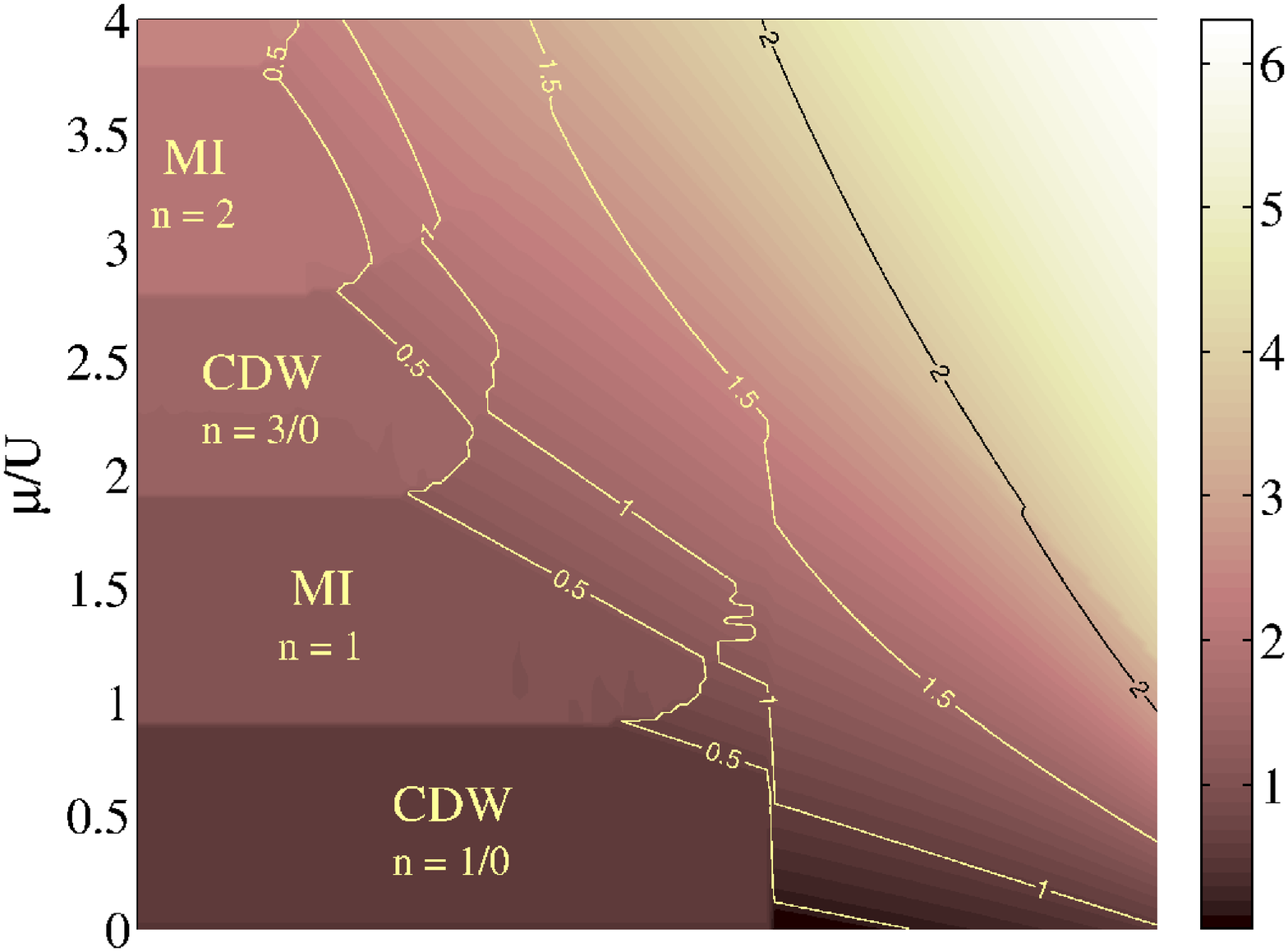}%
  \includegraphics[width=0.5\linewidth]{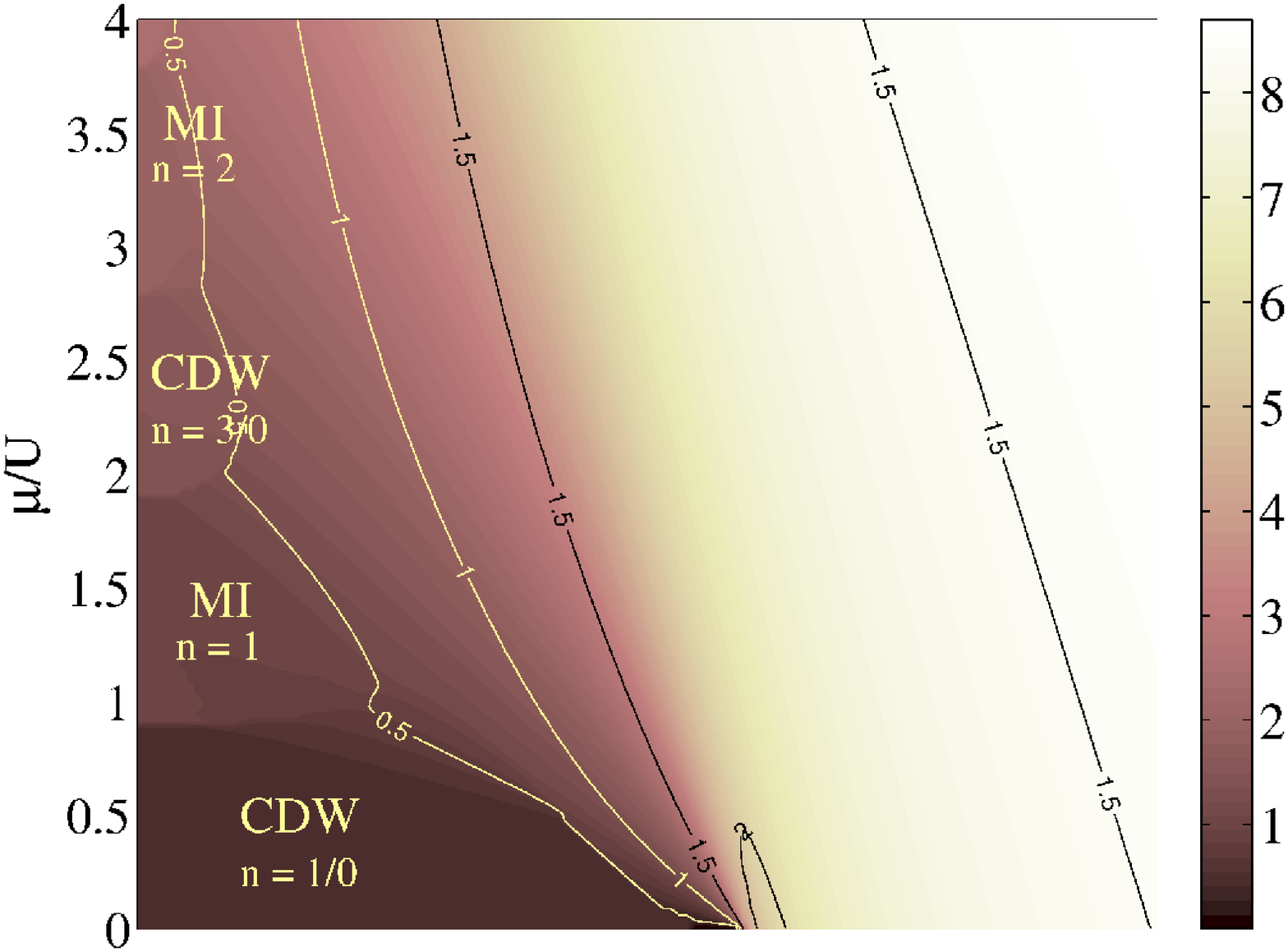}
  \includegraphics[width=0.5\linewidth]{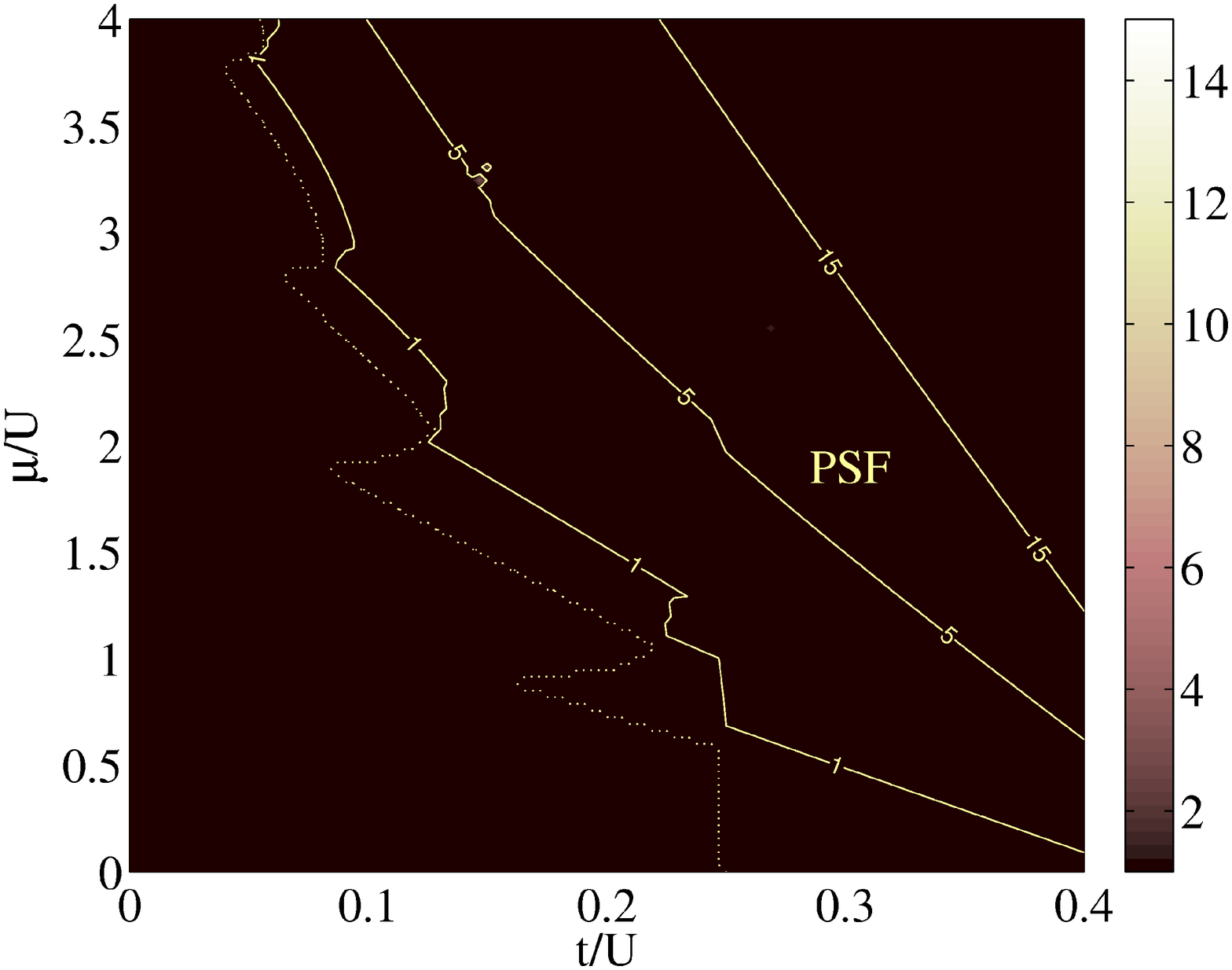}%
  \includegraphics[width=0.5\linewidth]{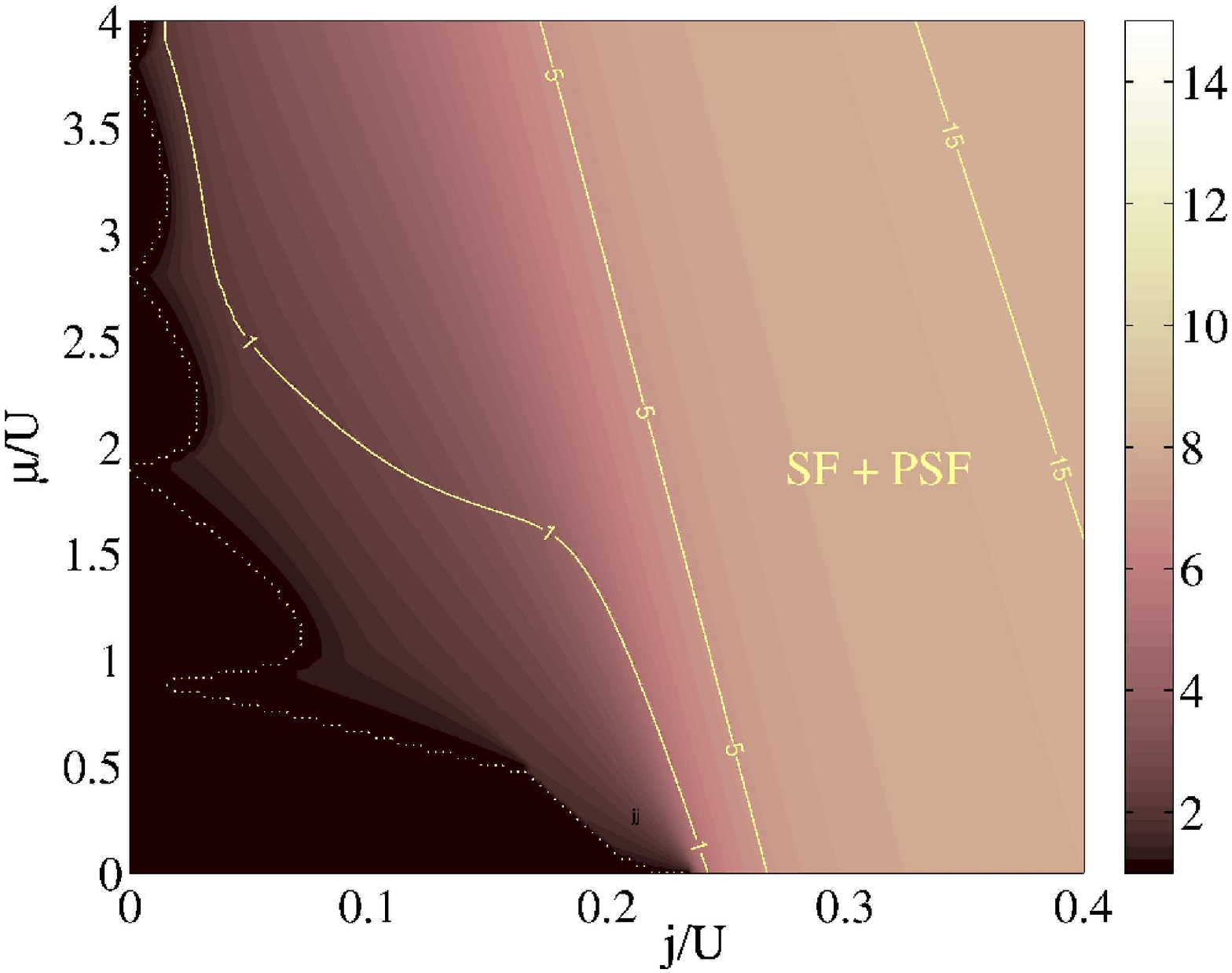}
  \caption{Phase diagram with a Gutzwiller ansatz for $U=1,V=-0.5.$
    We focus on (left) $j=0$ and (right) $t=0,$ separately. Upper
    plots show average site occupation $\langle n\rangle$ (grayscale),
    and number variance $\Delta n$ (contour). Lower plots show
    single-particle coherence $\langle a\rangle$ (grayscale)
    together with $\Delta a^2$ (contour).}
  \label{fig:GWminus}
\end{figure}

The first method that we use is a variational estimate of ground
state properties based on a product state \cite{krauth92}
\begin{equation}\label{eq:gutzwiller}
  |\psi_\mathrm{GW}\rangle =
  \prod_{i} \sum_{n_i} f_n^{(i)} \frac{1}{\sqrt{n_i!}}
  c_{i}^{\dagger n_i} \ket{0}.
\end{equation}
Minimizing the expectation value of the free energy $F = H - \mu N$
with respect to the variables $f_n,$ under the constrain of fixed
norm $\sum_n |f_n|^2 = 1,$ we will obtain the phase diagram in the
phase space of interactions and chemical potential $(U,V,j,t,\mu).$

In our study we have made several simplifications. First of all, we
assumed period-two translational invariance in the wavefunction,
using only two different sets of variational parameters,
$f_n^{(2i+1)}=f_n^1$ and $f_n^{(2i)}=f_n^0.$ In our experience, this
is enough to reproduce effects such as the CDW. Next, since $U\ge 0$
is required for the stability of the system, we have taken $U=1$ as
unit of energy. The limit $U=0$ is approximated by the limits $j,
t\gg 1$ in our plots. Finally, in order to determine the roles of
$j$ and $t,$ we have studied the cases $j=0$ and $t=0$ separately.
\begin{figure}
  \centering
  \includegraphics[width=0.5\linewidth]{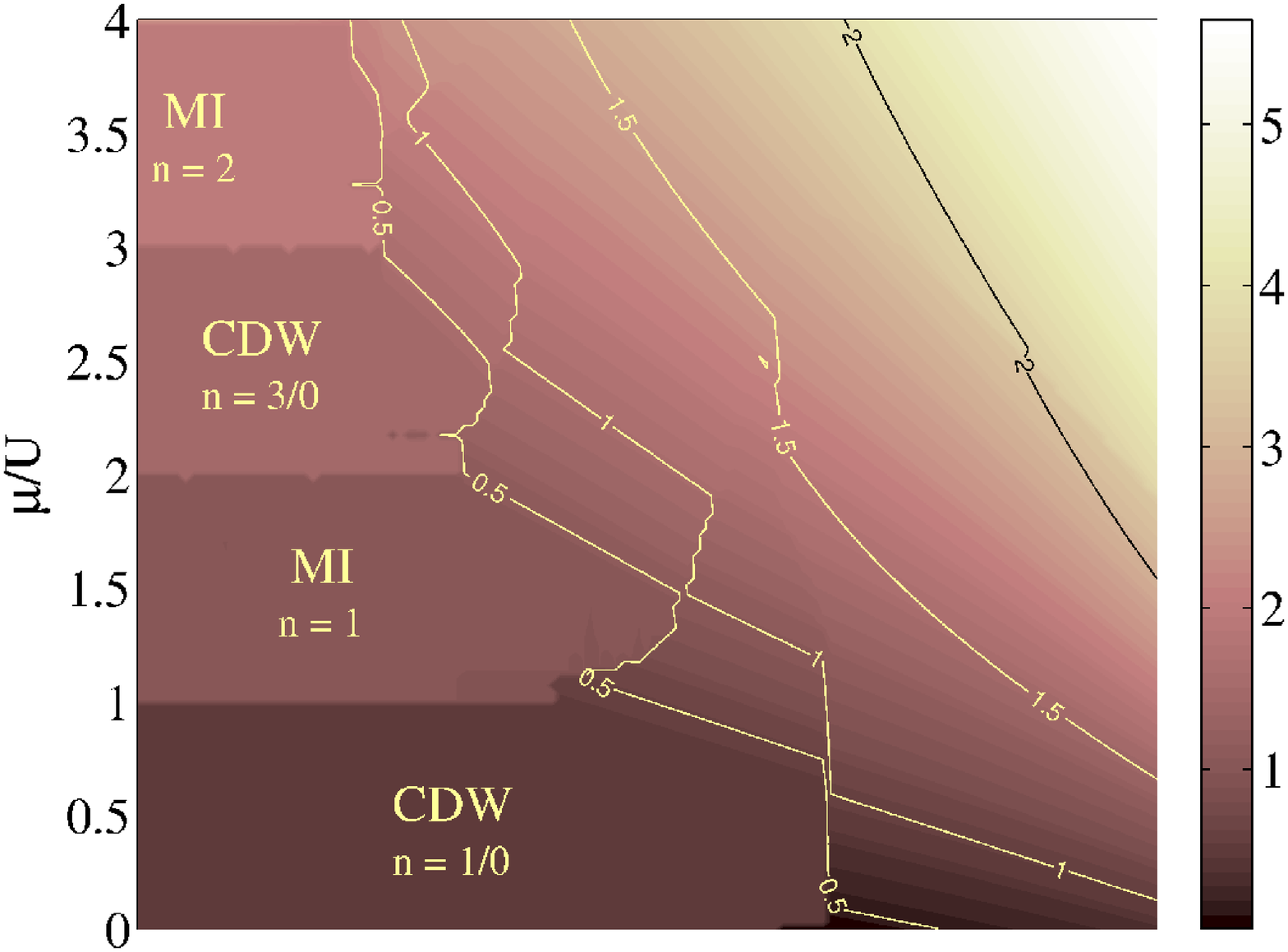}%
  \includegraphics[width=0.5\linewidth]{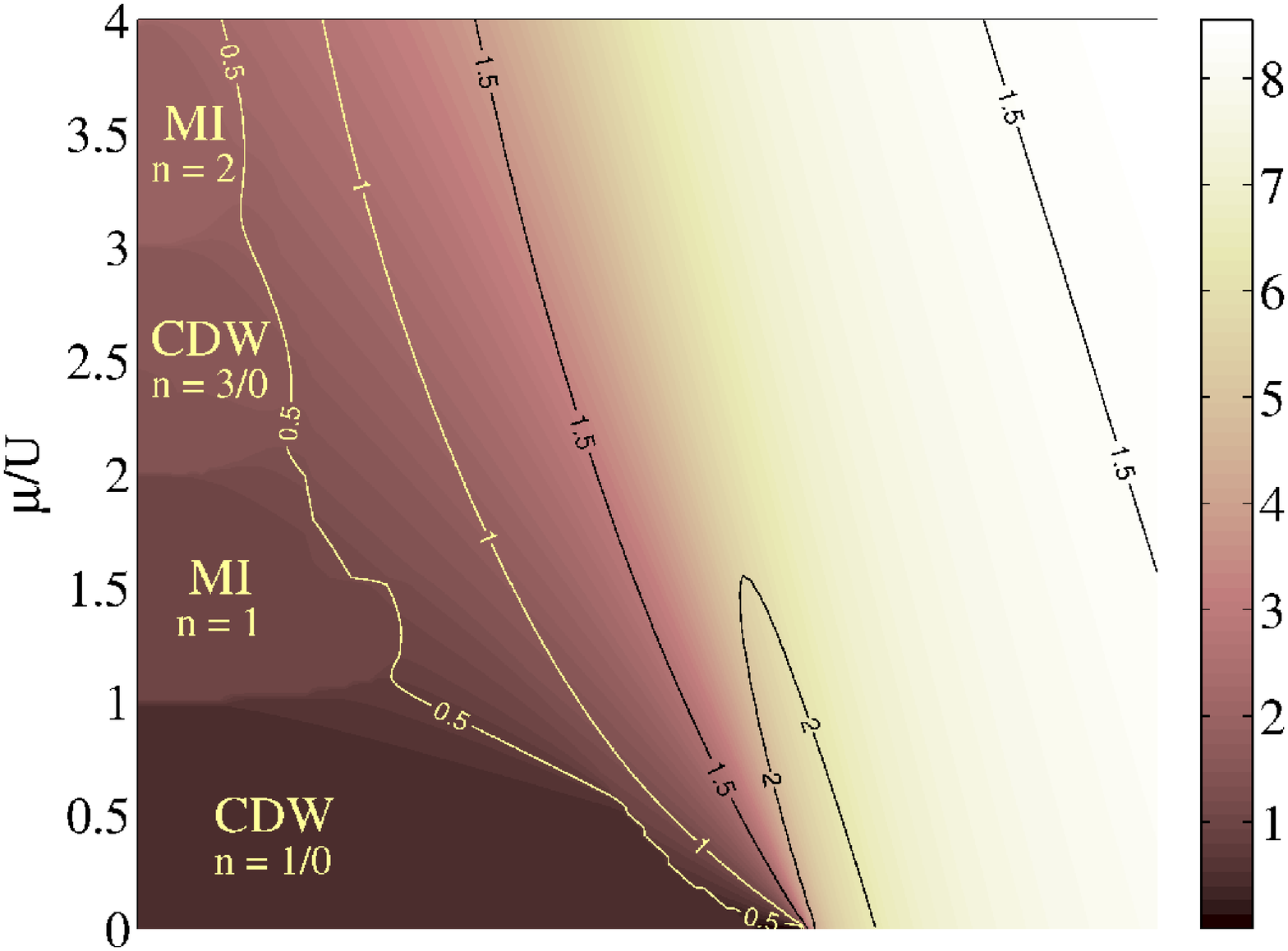}
  \includegraphics[width=0.5\linewidth]{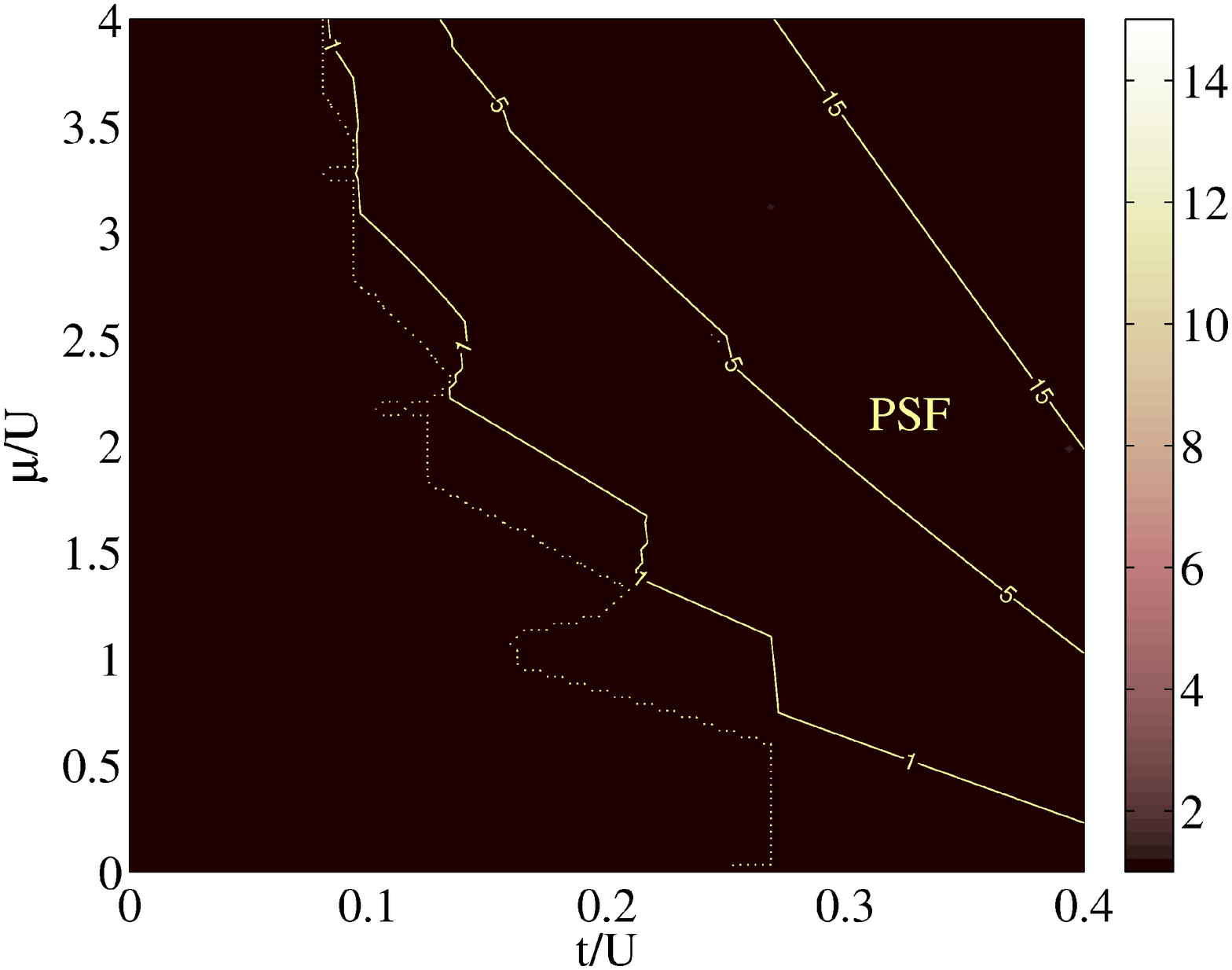}%
  \includegraphics[width=0.5\linewidth]{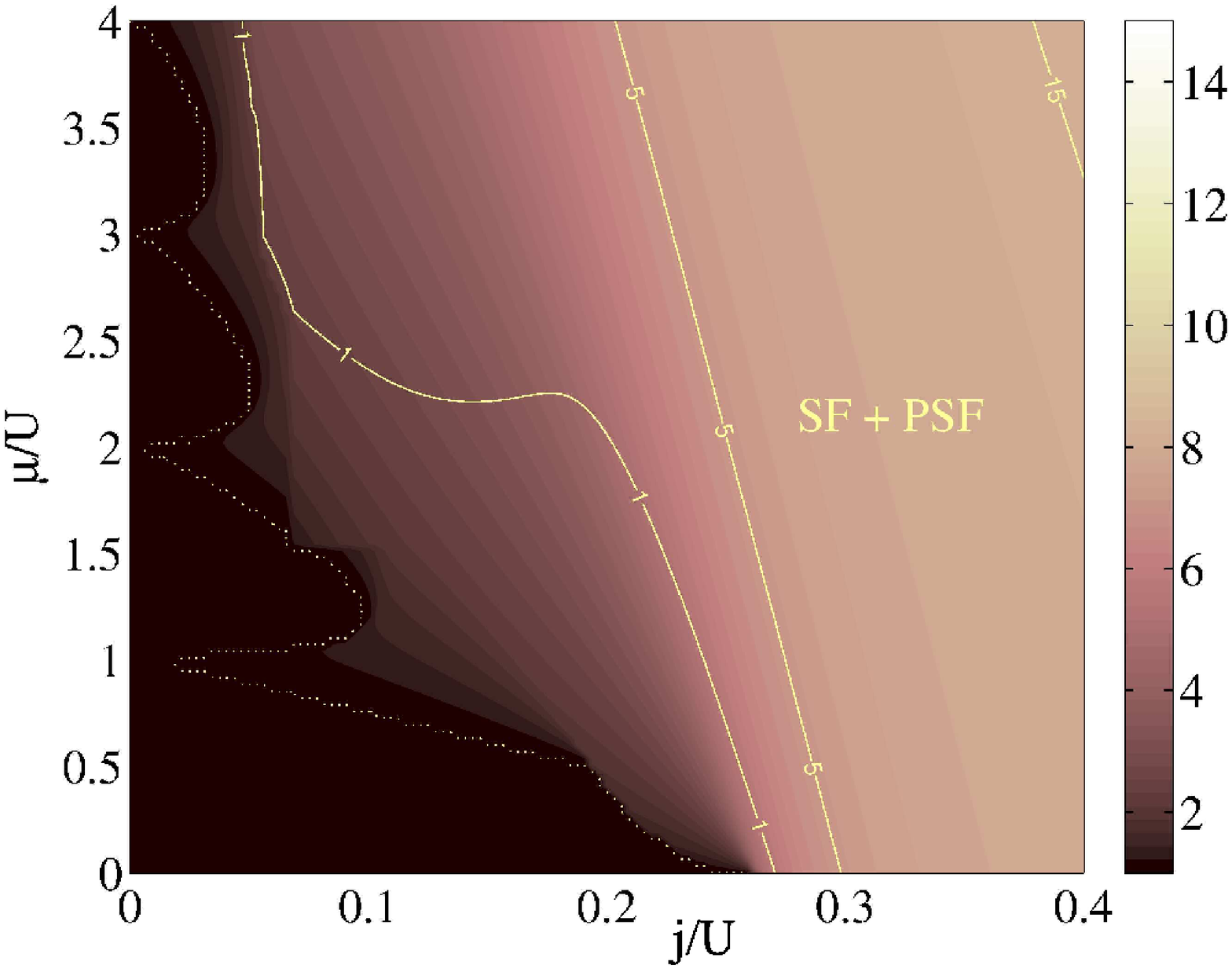}
  \caption{Phase diagram with a GW ansatz for $U=1,V=0.5.$ We focus on
    (left) $j=0$ and (right) $t=0,$ separately. Upper plots show average
    site occupation $\langle n\rangle$ (grayscale) and number variance
    $\Delta n$ (contour). Lower plots show single-particle coherence
    $\langle a\rangle$ (grayscale) together with $\Delta a^2$ (contour).}
  \label{fig:GWplus}
\end{figure}
The results are shown in Fig.~\ref{fig:GWminus} and
Fig.~\ref{fig:GWplus} for $V<0$ and $V>0,$ respectively.

The first interesting feature is that, as predicted by perturbation
theory, we have large lobes both with integer $1, 2, \ldots$ and
with fractional $1/0, 2/1, \ldots$ occupation numbers, forming
uniform Mott insulators and CDW, respectively. The insulators are
characterized by having a well defined number of particles per site,
and thus no number fluctuations $\Delta n^2 = \langle n^2\rangle -
\langle n\rangle^2 = 0.$ While the size of the lobes does not depend
dramatically on the sign of $V,$ these are significantly larger for
the pair hopping $t$ than for the correlated hopping $j,$ as already
seen with perturbation theory.

The boundary of the insulating areas marks a second order phase transition
to a superfluid regime, where we find number fluctuations $\Delta n\neq 0.$
In order to characterize these gapless phases we have computed the order
parameter of a single-particle condensate $\langle a\rangle,$ and two
quantities that we use to detect pairing. The first one is a two-particle
correlation that generalizes the order parameter of a BEC to the case of a
pair-BEC $\langle a^2\rangle.$ The second quantity $\Delta a^2 = |\langle
a^2\rangle - \langle a\rangle^2|$ is used to correct the previous value
eliminating the contribution that may come from a single-particle
condensate coexisting with the pair-BEC.

When $j=0$ we always find that $\langle a\rangle=0,$ even outside the
insulating lobes. This marks the absence of a single-particle BEC, which is
expected since we do not have single-particle hopping. On the other hand,
we now find long range coherence of the pairs and thus $\langle a^2\rangle
\neq 0$ all over the non-insulating area, which we identify with the
pair-superfluid regime.

The situation is slightly different for $t=0.$ The single-particle order
parameter $\langle a\rangle$ no longer vanishes in the superfluid area,
denoting the existence of single-particle coherence, but at the same time
we find that the two-particle correlations exceed the contribution from the
single-particle superfluid as $\Delta a^2 \neq 0,$ which we attribute to a
coexistence of both a single-particle and a pair-superfluid, or a state
with both features.

This picture does not change substantially when $V$ is positive or
negative. The only differences are in the insulating regions, where
the CDW is either due to the incommensurability of the particle
number $(V < 0)$ or really gives rise to the separation of particles
alternating holes and filled sites $(V > 0).$ However, in the
superfluid regime we find no significant changes and in particular
we see no breaking of the translational invariance or modulation of
the coherent phase.

\subsection{Matrix Product States: long range pair correlations}

The previous numerical simulations are very simple and cannot fully
capture the single particle and two-particle correlators. To
complete and verify the full picture we have searched the ground
states of the full Hamiltonian using the so called iTEBD algorithm,
which uses an infinite Matrix Product State ansatz together with
imaginary time evolution \cite{vidal07}.

\begin{figure}[t]
  \centering
  \includegraphics[width=0.5\linewidth]{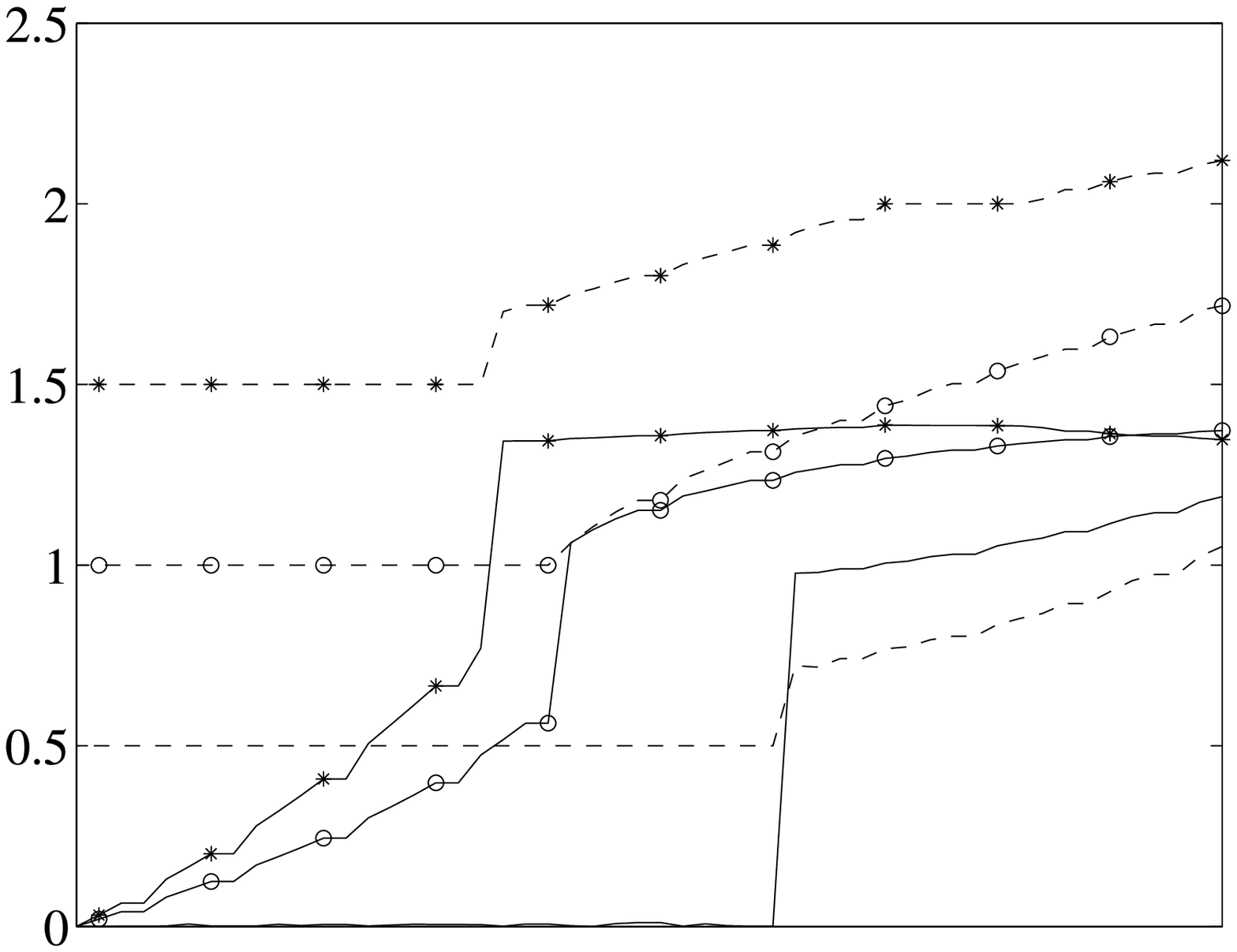}%
  \includegraphics[width=0.5\linewidth]{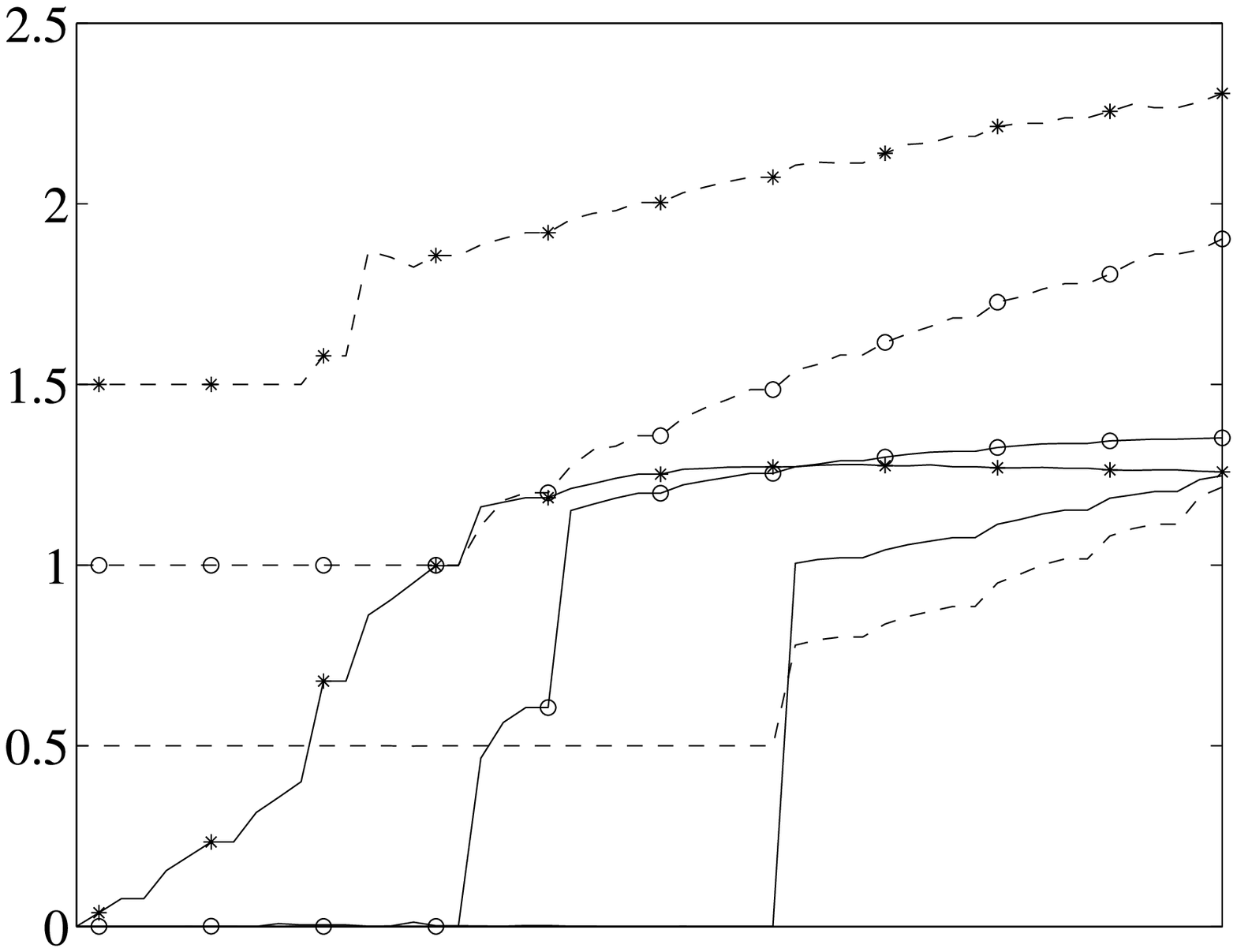}
  \includegraphics[width=0.5\linewidth]{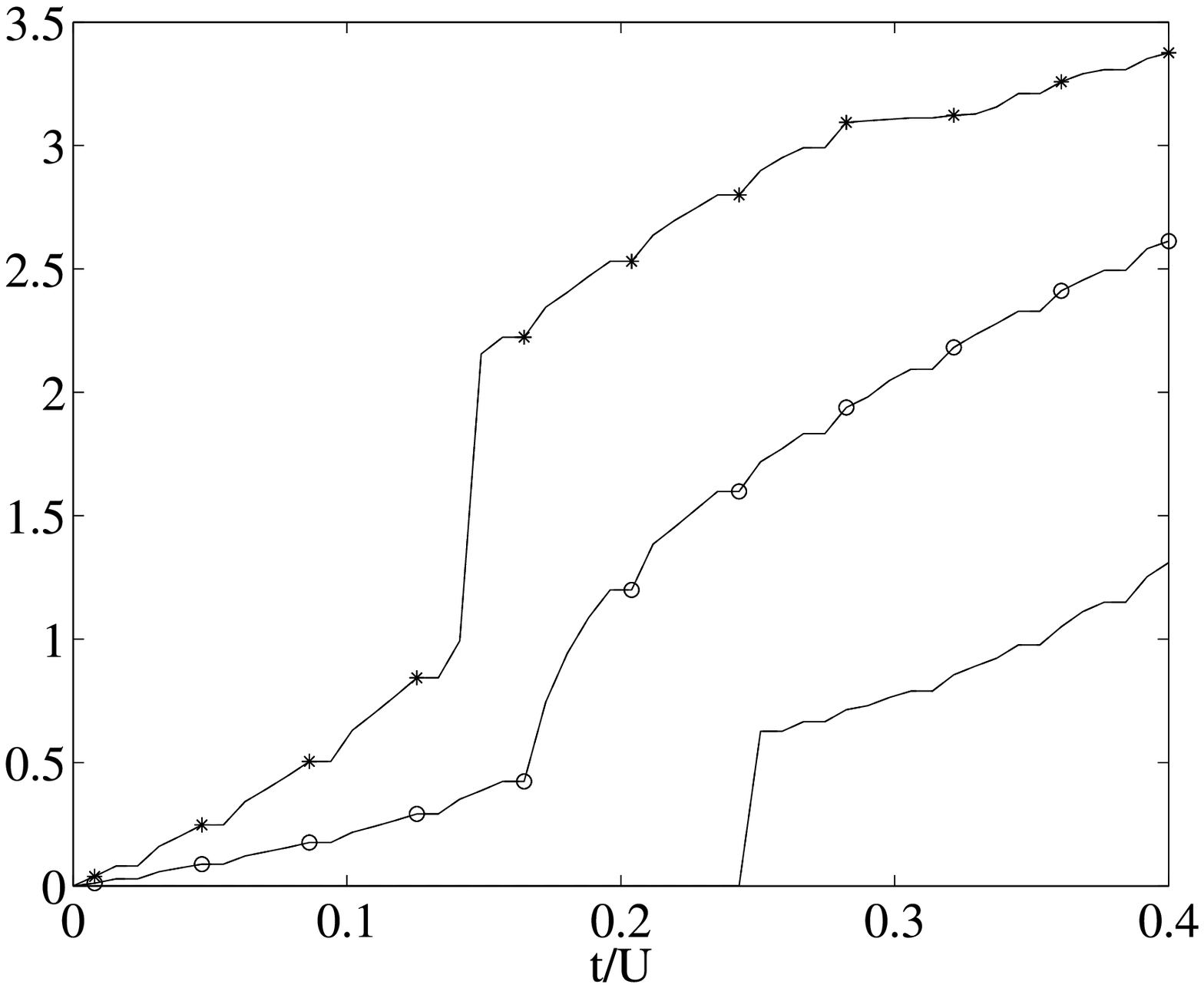}%
  \includegraphics[width=0.5\linewidth]{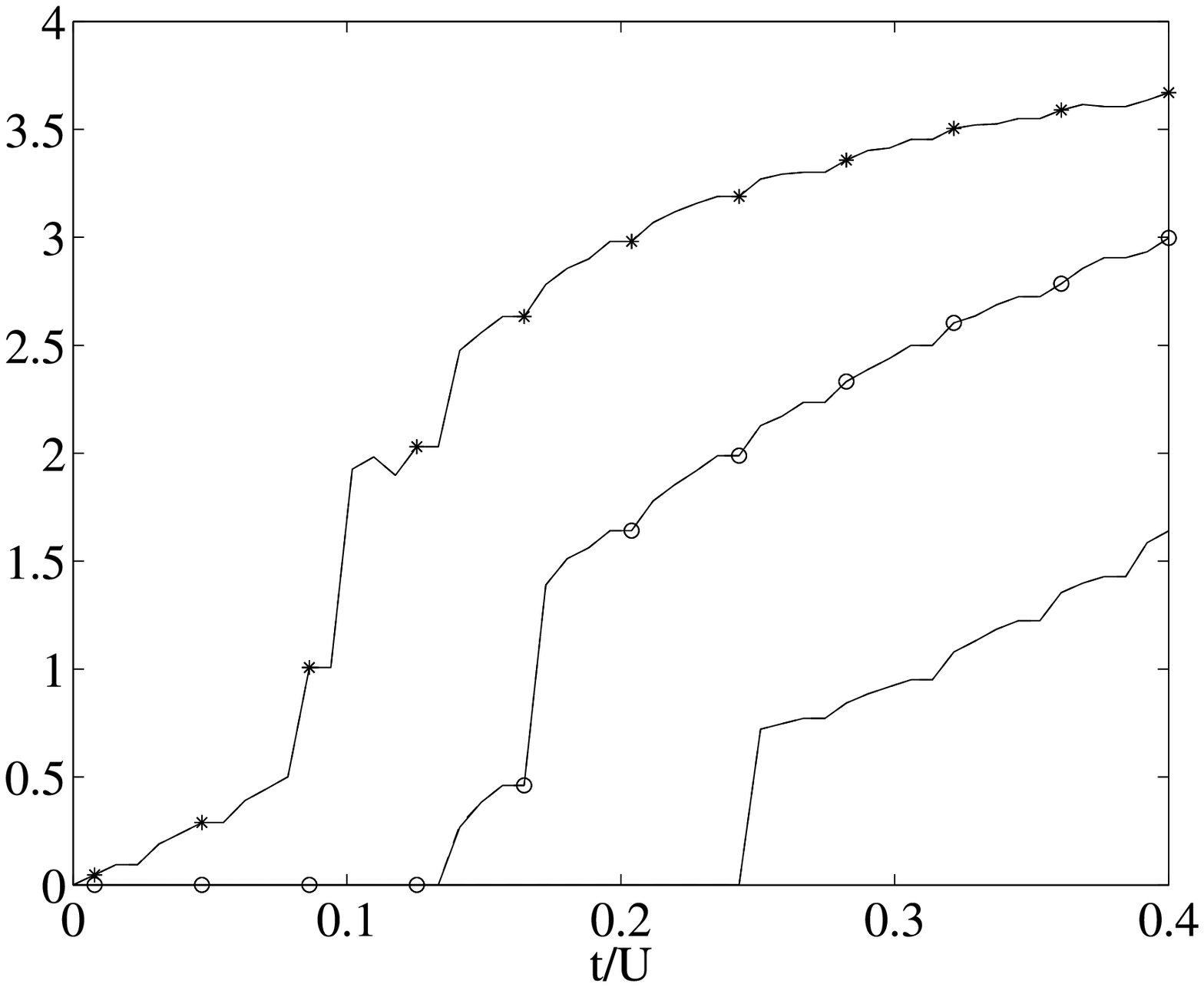}
  \caption{Numerical simulations with the iTEBD algorithm. On the left we
    plot the results for $j=0,U=1,V=-0.05$ and on the right for $V=0.05,$
    as a function of two-particle tunneling $t.$ The upper row shows the
    density (dashed) and particle number fluctuations (solid), while the
    lower row shows $(\Delta a)^2$ (solid) and two-particle coherence
    (dashed), which overlap indicating $\langle a\rangle = 0.$ For each set
    of operators we show three lines, for $\mu/U=0.5,~1.5$ and $2.5$
    corresponding to plain line, circle and star, respectively.}
  \label{fig:mps}
\end{figure}

Roughly, this ansatz is based on an infinite contraction of tensors
that approximates the wavefunction of a translational invariant
system in the limit of infinite size. Adapting the ansatz to our
problem we write it as
\begin{eqnarray}
  \left | \psi \right \rangle \sim \prod_{k\in\mathbb{Z}}
  \Gamma^{o}_{\alpha_{2k+1} \alpha_{2k+2}}(n_{2k+1})
  \lambda^{o}_{\alpha_{2k+2}}\Gamma^{e}(n_{2k+2})_{\alpha_{2k+2} \alpha_{2k+3}}
  \lambda^{e}_{\alpha_{2k+3}}\times&&\nonumber\\
 \quad\quad \quad\quad\times
  \frac{1}{\sqrt{n_{2k+1}! n_{2k+2}!}}a^{\dagger n_{2k+1}}_{2k+1}a^{\dagger n_{2k+2}}_{2k+2}
    \left|\mathrm{vac}\right\rangle.&&
\end{eqnarray}
Here the $\Gamma^{o}$ and $\Gamma^{e}$ are matrices that depend on the
state of the odd and even sites they represent, a dependence which is
signaled by the $n_{2k+1}$ and $n_{2k+2}$ in the previous equation. These
matrices are contracted with one-dimensional vectors of positive weights
$\lambda^{e,o}_\alpha \geq 0,$ which are related to the coefficients of the
Schmidt decomposition. This variational ansatz is known to work well for
states with fast decaying correlations, but it also gives a good
qualitative description of the critical phases.

In order to optimize the iTEBD wavefunction we performed an approximate
imaginary time evolution using a Trotter decomposition and local updates of
the associated tensors, as described in Ref.~\cite{orus08}. Using the canonical
forms for these tensors it is also straightforward to compute expectation
values for different operators acting either on neighboring or separated
sites.

In Fig.~\ref{fig:mps} we plot the most relevant results for three cuts
across the phase diagram, $\mu=0.5,~1.5$ and $2.5,$ so that each line
crosses both an insulating plateau and the superfluid region. We have used
small tensor sizes from $D=16$ up to $64,$ a value limited by the need of
using large cutoffs for the site populations ($n_{max}=8$). As shown in the
figures, when $j=0$ the single-particle correlator is zero for distinct
sites, and we are left only with two-particle correlations. In the MI case
the pair correlations between neighboring sites decrease very quickly,
while in the superfluid regime we see a critical behavior
\begin{equation}
  C^2_\Delta \propto \frac{1}{\Delta^\alpha},
\end{equation}
with an exponent that varies between $\alpha=0.5$ and $\alpha=0.6,$
depending on the simulation parameters.

\begin{figure}
  \centering
  \includegraphics[width=0.5\linewidth]{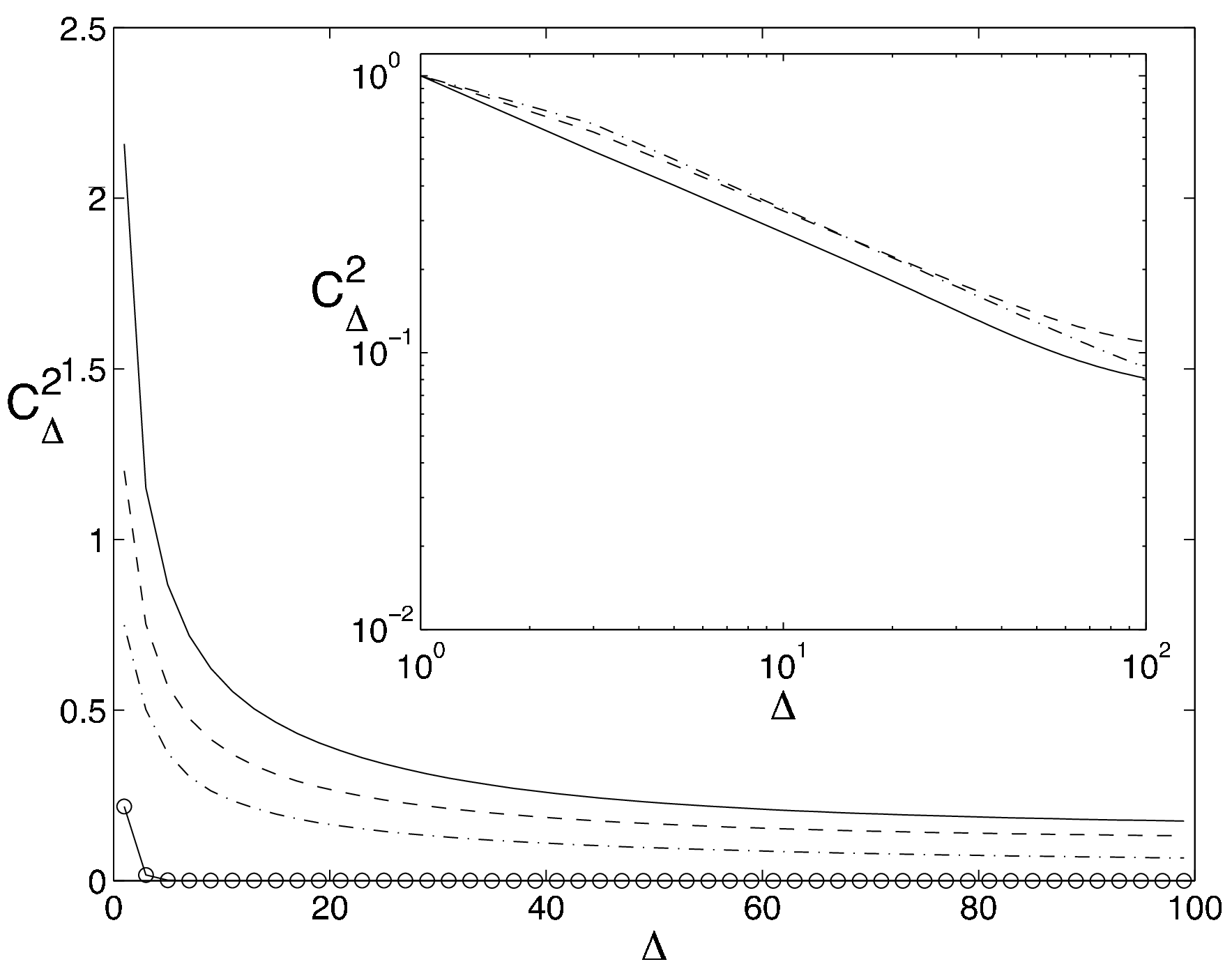}
  \caption{Correlator $C^2_\Delta = \langle{a^{\dagger 2}_{i+\Delta}a^2_i}
    \rangle$ vs. site separation, for $(t,\mu)$ $=$ $(0.1,1.5),$
    $(0.3,0.5),$ $(0.2,1.5),$ $(0.1,2.5)$, from top to bottom. The line with
    $t=0.1,\mu=1.5$ (solid, circles) corresponds to a MI state and
    correlations decay to zero on the third site. The remaining lines have
    been rescaled and plotted in log-log scale in the inset, which shows
    that the two-particle correlator decays approximately as
    $\Delta^{-\alpha}$ with a power $\alpha \sim 0.5-0.6.$}
  \label{fig:mps2}
\end{figure}

\section{Detection}\label{sec:detection}

There are many ways of differentiating the phases we have found,
each one having its own degree of difficulty. The simplest and best
established detection methods are related to the insulating regimes.
These phases, which involve both MI and CDW, are characterized by
having a well defined number of particles at each lattice site, the
lack of coherence and an energy gap that separates the insulator
from other excitations.

The energy gap in these insulators may be probed either by static or
spectroscopic means as it has been done in experiments
\cite{greiner02,stoeferle04}, determining that indeed the system is
insulating. Second, the lack of coherence will translate into
featureless time-of-flight images, having no interference fringes
\cite{greiner02} at all. Even though there will be no fringes, the
measured density will be affected by quantum noise. The analysis of
the noise correlation will show peaks at certain momenta
\cite{altman04,foelling05} that depend on the periodicity of the
state, so that the number of peaks in the CDW phase will be twice
those of the MI. In case of having access to the lattice sites, as in
the experiments with electron microscopy \cite{scan08}, or in future
experiments with large aperture microscopic objectives that can
collect the fluorescence of individual lattice sites
\cite{weiss07,karski09}, the discrimination between the MI and the CDW
should be even easier, since in one case we have a uniform density and
in the other a periodic distribution of atoms.

When the system enters a superfluid phase, it becomes a perfect
``conductor'' with a gapless excitation spectrum. The lack of an energy
gap, should be evident in the spectroscopic experiments suggested
before. However, we are not only interested in the superfluid nature, but
rather in the fact that this quantum phase is strongly paired. More
precisely, we have found that for $j=0$ the single-particle coherence is
small or zero, and that the two-particle correlator decays slowly
\begin{eqnarray}
  C^1_\Delta &=& \langle a^\dagger_i a_{i+\Delta}\rangle
  \sim \bar n \delta_{\Delta 0}\\
  C^2_\Delta &=& \langle a^{\dagger 2}_ia^2_{i+\Delta}\rangle
  \sim 1/\Delta^\alpha.
\end{eqnarray}
The first equation implies that the time of flight images will
reveal no interference fringes and will exhibit noise correlations
which will be similar to the MI. In order to probe $C^2_\Delta$ and
confirm the pairing of the particles, we suggest to use Raman
photo-association to build molecules out of pairs of atoms
\cite{stwalley98,wynar00}. For an efficient conversion, it would be
best to perform an adiabatic passage from the free atoms to the
bound regime. As described in \cite{ripoll06}, we expect a mapping
that goes from $\left|2n\right\rangle \rightarrow
\left|n\right\rangle,$ where $2n$ is the number of bosons and $n$
the number of molecules. More precisely, we expect the $a^{\dagger
2}$ operator to be mapped into $m^\dagger,$ so that the pair
coherence of the original atoms translates into the equivalent of
$C^1_\Delta$ for the molecules. This order should reveal as an
interference pattern in time-of-flight images of the molecules.

Finally, in cases with $j\neq 0,$ we have found the coexistence of
single-particle and two-particle coherences. This translates also
into the coexistence of interference fringes with nonzero pair
correlators.

\section{Conclusions}\label{sec:summary}

Summing up, we have suggested a family of experiments with cold
atoms that would produce correlated hopping of bosons. The mechanism
for the correlated hopping is an asymmetry in the contact
interaction between atoms. This asymmetry is exploited by trapping
the atoms in dressed states, a configuration that gives rise to
transport induced by collisions.

The main result of this paper is that there is a huge variety of
interaction asymmetries that will give rise to long range pair
correlations via interaction--induced transport.  Formally, in the
resulting effective models we recognize two dynamical behaviors. If we
have a nonzero asymmetry in the interspecies interactions,
\begin{equation*}
  t \propto 2g_{\uparrow\downarrow} - g_{\uparrow\uparrow} -
  g_{\downarrow\downarrow} \neq 0,
\end{equation*}
the Hamiltonian will exhibit pair hopping, while an asymmetry in the
intra-species scattering lengths
\begin{equation*}
  j \propto g_{\uparrow\uparrow} - g_{\downarrow\downarrow} \neq 0,
\end{equation*}
gives rise to correlated hopping.  However, we have given enough
evidence that \textbf{both} Hamiltonian terms give rise to a novel
quantum phase which we call pair superfluid. This phase is
characterized by a gapless spectrum with a finite sound speed, zero
single-particle correlations and long range pair coherence. All
quantum phases are connected by second order quantum phase
transitions. These phases can be produced and identified using
variations of current experiments \cite{anderlini06, sebbystrabley06,
  foelling05}. The nonperturbative nature of the effect should help in
that respect.

Our ideas are not restricted to one dimension. It is possible to engineer
also a two--body hopping using two--dimensional lattice potentials. Again,
the basic ingredients would be atoms with an asymmetric interaction and an
optical lattice that traps two states, $\ket{+}$ and $\ket{-}$ with a
relative displacement. Both in the one-- and two--dimensional cases it is a
valid question to ask whether the coupling between different trapped
states, $\ket{\pm},$ can excite also transitions to higher bands, processes
that have not been considered in the paper. Our answer here is no. There
are only two sources of coupling to higher energy bands. One is the
interaction, but we are already assuming that the interaction energies are
much smaller than the band separation. Following the notation from
Ref.~\cite{jaksch98}, we have the constraint that the interaction energy
should be smaller than the energy separation to the first excited state in
a well of the periodic potential, $\bar{n}^2U \ll \hbar\bar\nu\bar{n}$ the
same requisite as for ordinary Bose--Hubbard models \cite{greiner02}. The
other source of coupling to higher bands would be single--particle
hopping. However, unlike \cite{jaksch98}, here we are assuming that these
terms are strongly suppressed compared with the interaction. In other
words, realizing the models that we suggest in this paper, for realistic
densities, $\bar n=2,$ and simple potentials, imposes no further constraint
in current experiments.

Finally, let us remark that transport--inducing collisons may be
implemented using other kinds of spin-dependent interactions. For instance,
correlated hopping appears naturally in state-dependent lattices loaded
with spinor atoms, because their interactions can change the hyperfine
state of the atoms while preserving total angular momentum \cite{widera06}.
\newline

We would like to thank Miguel Angel Mart{\'\i}n-Delgado for useful
discussions. M.E. acknowledges support from the CONQUEST project.
J.J.G.R acknowledges financial support from the Ramon y Cajal
Program of the Spanish M.E.C., from U.S. NSF Grant No. PHY05-51164
and from the spanish projects FIS2006-04885 and CAM-UCM/910758.

\appendix
\section{Derivation of the model in superlattices}
\label{sec:derivation}

As discussed before, the main idea behind atomic correlated hopping
is to trap atoms whose interaction allows them to change their
state. In this section we provide one possible implementation of
this idea using state dependent superlattices that trap dressed
states.

\subsection{Dressed states trapping}
\label{sec:trapping}

Our starting point is the setup in Fig.~\ref{fig:superlattice}a,
which is itself taken from Ref.~\cite{garciaripoll07}. It consists
on an optical lattice trapping atoms in states $\ket{\uparrow}$ and
$\ket{\downarrow},$ together with a Raman coupling between these
states. Mathematically, this configuration is described by the
single-particle Hamiltonian
\begin{eqnarray}
  H_{\mathrm{trap}} &=& V_0~\sin(kx)^2 \left( \ket{\uparrow}\bra{\uparrow}
    + \ket{\downarrow}\bra{\downarrow} \right ) \nonumber\\ & & + \Omega~
    \sin(kx) \left(\ket{\uparrow}\bra{\downarrow} +
    \ket{\downarrow}\bra{\uparrow}\right).
\end{eqnarray}
By moving to the basis of dressed states $\ket{\pm} =
\frac{1}{\sqrt{2}} \left(\ket{\uparrow} \pm \ket{\downarrow}
\right),$ we find that the trapping is effectively equivalent to two
superlattices with a relative displacement as in
Fig.~\ref{fig:superlattice}b
\begin{eqnarray}
  H_{\mathrm{trap}} &=&
  \left(V_0 \sin(kx)^2+ \Omega \sin(kx)\right) \ket{+}\bra{+} \nonumber
  \\ & & + \left(V_0 \sin(kx)^2 - \Omega \sin(kx)\right) \ket{-}\bra{-}.
\end{eqnarray}
Under appropriate circumstances, discussed in \cite{garciaripoll07},
we will find that each superlattice site has a unique ground state,
energetically well differentiated from the next excited state, and
which consists of a symmetric wavefunction spanning both lattice
wells. If this is the case and if all energy scales, such as the
interaction and the hopping, are small compared to the separation
between Bloch bands, we can expand the bosonic field operators
describing the atoms in terms of these localized wavefunctions
\begin{eqnarray}
  \psi_+(x) = \sum_i c_{2i}~ W(x - 2il),\label{eq:tight-binding}\\
  \psi_-(x) = \sum_i c_{2i+1}~ W(x - (2i+1)l),\nonumber
\end{eqnarray}
where $2l=2\pi/k$ is the superlattice period, $c_{j}$ are bosonic
operators that, for $j$ even (odd), annihilate an atom in state
$\ket{+}$ ($\ket{-}$) in the $j$-th superlattice cell
\begin{eqnarray}
  c_{2i} = \frac{1}{\sqrt{2}} (a_{2i,+} + a_{2i+1,+}),\label{eq:deloc-ops}\\
  c_{2i+1} = \frac{1}{\sqrt{2}} (a_{2i+1,-} + a_{2i+2,-}),\nonumber
\end{eqnarray}
and the localized wavefunctions $W(x)$ are a superposition of the
Wannier functions $w(x')$ of the underlying lattice
\begin{equation}\label{eq:W}
W(x-2il) = \frac{1}{\sqrt{2}} \left[w(x-2il) + w(x-(2i+1)l)\right].
\end{equation}

\subsection{State-changing collisions}
\label{sec:collisions}

We will now express the interaction (\ref{eq:interaction2}) in the
basis of dressed states. We proceed using the change of variables in
Eq.~\ref{eq:dressed} to find the expression of the densities
\begin{eqnarray}
  \rho_\uparrow(x) &=& \frac{1}{2}(\rho_{+}  + \rho_{-} +
  \psi_+^\dagger \psi_- + \psi_-^\dagger \psi_+),\\
  \rho_\downarrow(x) &=& \frac{1}{2}(\rho_{+}  + \rho_{-} -
  \psi_+^\dagger \psi_- - \psi_-^\dagger \psi_+).
\end{eqnarray}
The first obvious conclusion is that the total density is
independent of the basis on which it is written,
\begin{equation}\label{eq:total-density}
  \rho(x) = \rho_\uparrow(x) + \rho_\downarrow(x) = \rho_+(x) + \rho_-(x).
\end{equation}
Hence, the term of $g_0$ is insensitive to the state of the atoms.
On the other hand, the asymmetric terms are not so simple. The $g_1$
interaction, which is proportional to the product of densities
\begin{eqnarray}
  :\rho_\uparrow \rho_\downarrow: &=& \frac{1}{4} :(\rho_+ + \rho_-)^2:
  - \frac{1}{4}: (\psi_+^\dagger \psi_- + \psi_-^\dagger \psi_+)^2 :
  \nonumber\\
  &=& \frac{1}{4} :(\rho_+ + \rho_-)^2: - \frac{1}{2} \rho_+\rho_- -
  \frac{1}{4}(\psi_+^{\dagger 2}\psi_-^2 + H.c)
  \nonumber\\
  &=& \frac{1}{4} :\rho_+^2 + \rho_-^2: -
  \frac{1}{4}(\psi_+^{\dagger 2}\psi_-^2 + H.c),\label{eq:asym}
\end{eqnarray}
gives rise to a scattering that changes the state of interacting
atoms from $\ket{-}$ to $\ket{+}$ and viceversa, as in
Fig.~\ref{fig:collisions}a. The term of $g_2$ has a lightly
different effect,
\begin{equation}\label{eq:imbalance}
  :\rho_{\uparrow}(x)^2-\rho_{\downarrow}(x)^2: =
  :\rho(x) \left[\psi^\dagger_+(x)\psi_-(x) +
  \psi^\dagger_-(x)\psi_+(x)\right]:
\end{equation}
it gives rise to processes where one atom changes its state
influenced by the surrounding environment. In the following
subsections we will see what happens to the interaction terms
(\ref{eq:total-density}), (\ref{eq:asym}) and (\ref{eq:imbalance}),
when the atoms are confined in a lattice.

\subsection{Final model}
\label{sec:final-model}

In this section we will put the previous results of this appendix
together. We will take the tight-binding expansion of the field
operators (\ref{eq:tight-binding}) and use it together with
Eqs.~(\ref{eq:total-density}), (\ref{eq:asym}) and
(\ref{eq:imbalance}) to expand the interaction Hamiltonian
(\ref{eq:interaction2}). For convenience, we will rename the bosonic
operators as
\begin{equation}
  c_{2k} = a_{k+}~~\mathrm{and}~~c_{2k+1} = a_{k-}
\end{equation}
according to the position at which their Wannier functions are
centered (see Fig.~\ref{fig:superlattice}c). Along the derivation,
one obtains many integrals of ground state wavefunctions
\begin{equation}
  C_{k,m} = \int |W(x-kl)|^2 |W(x-ml)|^2 dx.
\end{equation}
We will only keep those integrals with a separation smaller than a
superlattice period. Taking Eq. (\ref{eq:W}), the expression for the
superlattice localized states, one obtains
\begin{eqnarray}
  C_{k,k} &=& \int |W(x)|^4 dx \simeq \frac{1}{2} \int |w(x)|^4 dx,\\
  C_{k,k\pm1} &=& \int |W(x)|^2|W(x-l)|^2 dx \simeq
  \frac{1}{4} \int |w(x)|^4 dx,
\end{eqnarray}
where $w(x)$ are the Wannier wavefunctions of the underlying
sublattice. Using these tools, the symmetric interaction term becomes
\begin{eqnarray}
\fl
~~~~~\frac{g_0}{2} \int dx :(\rho_\uparrow(x) + \rho_\downarrow(x))^2: =\nonumber\\
= \frac{g_0}{2} \sum_k^{N/2} :n_{2k}^2 C_{2k,2k} + n_{2k+1}^2
C_{2k+1,2k+1} + 2n_{2k}n_{2k+1} C_{2k,2k+1}: \nonumber\\
= \frac{g_0}{4}\int dx~|w(x)|^4~ \sum_k^N :n_k^2 + n_k n_{k+1}:
\end{eqnarray}
For the asymmetric parts, we first use Eq.~(\ref{eq:asym}) obtaining
\begin{eqnarray}
\fl
~~~~~g_1\int dx :\rho_\uparrow(x) \rho_\downarrow(x): = \nonumber\\
\frac{g_1}{8}\int dx~|w(x)|^4~ \sum_k^N \left[:n_k^2: -~
\frac{1}{2}\left(c_{k+1}^{\dagger 2}c_k^2 + c_k^{\dagger
2}c_{k+1}^2\right) \right]
\end{eqnarray}
and then finally the more complicated Eq.~(\ref{eq:imbalance})
\begin{eqnarray}
\fl ~~~~~\frac{g_2}{2}\int dx :\rho_\uparrow(x)^2 -
\rho_\downarrow(x)^2: =
\nonumber\\
\frac{g_2}{8}\int dx~|w(x)|^4 \sum_k^{N} : n_{k}(c_{k}^\dagger
c_{k-1} + c_{k-1}^\dagger c_{k} + c_{k}^\dagger c_{k+1} +
c_{k+1}^\dagger c_{k}) :
\end{eqnarray}
Introducing constants that parameterize the on-site interactions and
the strength of the underlying lattice (\ref{int-constants}), our
final Hamiltonian looks as follows
\begin{eqnarray}
H & = & \frac{2U_0+U_1}{8}\sum_k :n_k^2: - \frac{U_0}{8} \sum_k :n_kn_{k+1}: -
    \frac{U_1}{16} \sum_k (c^{\dagger 2}_{k+1}c^{2}_{k} + \mathrm{H.c.})
    \nonumber\\
    & & - \frac{U_2}{8} \sum_k \left[(n_k - 1) c^\dagger_k(c_{k-1}+ c_{k+1}) +
    \mathrm{H.c.}\right].
\end{eqnarray}
Completing terms and replacing the sum over $k$ with a sum over
nearest neighbors, we arrive at the desired model (\ref{eq:model})
with the parametrization given already in Eq.~(\ref{eq:paramet}).

\section*{References}


\end{document}